\documentclass[floats,aps,showpacs]{revtex4}
\usepackage{amssymb}
\usepackage{amsmath}
\usepackage{epsfig}
\usepackage{graphicx}

\newcommand{\be}{\begin{equation}}
\newcommand{\ee}{\end{equation}}
\newcommand{\bea}{\begin{eqnarray}}
\newcommand{\eea}{\end{eqnarray}}

\newcommand{\p}{\partial}
\newcommand{\s}{\sigma}

\newcommand{\la}{\langle}
\newcommand{\ra}{\rangle}

\newcommand{\ri}{\mbox{i}}

\newcommand{\ua}{\uparrow}
\newcommand{\da}{\downarrow}
\def\nn{\nonumber\\}
\def\fr#1{(\ref{#1})}

\def\nn{\nonumber\\} 
\def\up{\uparrow}
\def\down{\downarrow}
\begin{document}

\title{Finite wave vector pairing in doped two-leg ladders}

\author{N. J. Robinson$^1$, F. H. L. Essler$^1$, E. Jeckelmann$^2$ and A. M. Tsvelik$^3$}
\affiliation{$^1$
The Rudolf Peierls Centre for Theoretical Physics, Oxford
University, Oxford OX1 3NP, UK\\
$^2$ Institute for Theoretical Physics, Leibniz Universitaet Hannover,
Appelstrasse 2, D-30167 Hannover, Germany \\
$^3$ Department of  Condensed Matter Physics and
Materials Science, Brookhaven National Laboratory, Upton, NY
11973-5000, USA} 
\date{\today}

\begin{abstract}
We consider the effects of Umklapp processes in doped two-leg
fermionic ladders. These may emerge either at special band fillings or
as a result of the presence of external periodic potentials. We show
that such Umklapp processes can lead to profound changes of physical
properties and in particular stabilize pair-density wave phases. 

\end{abstract}

\maketitle

\sloppy
\section{Introduction}
\label{sec:Intro}

As is well illustrated by the  example of the one-dimensional Hubbard
model \cite{book}, Umklapp processes in strongly correlated systems
may lead to a profound restructuring of the ground state. Indeed, at
half filling when the Fermi wave vector is such that $4k_F = 2\pi$,
Umklapp scattering processes connect opposite Fermi points and open a
spectral gap for single-particle excitations. In a similar way
Umklapp processes in undoped two-leg fermionic ladders are known to
generate a variety of insulating states
\cite{balents,O(8),tsuchiizu,fradkin}. In both cases these 
Umklapp processes become relevant at the particular density of one
electron per site, independently of the details of the
interactions. In multi-band systems such as the 2-leg ladder there are
other kinds of Umklapp processes that can connect Fermi points at
certain other band fillings, which generally depend on the microscopic
details of both the band structure and the interactions.

One example where such processes may play a crucial role is the
``telephone number compound'' ${\rm Sr_{14-x}Ca_xCu_{24} O_{41}}$
\cite{SrCaCuO1,SrCaCuO2}. X-ray scattering techniques have established
the presence of a standing wave in the hole density without a
significant lattice distortion in this material \cite{SrCaCuO1}. The
simplest explanation for these findings is a crystalline state of
pairs of holes 
\cite{PDW,Poilblanc}. The physical origin of the hole crystal is
likely to be the long-ranged Coulomb interaction between
ladders. Treating this interladder Coulomb interaction in a
mean-field approximation leads to a model of decoupled ladders subject
to a (self-consistent) periodic potential \cite{Poilblanc}.
The latter introduces Umklapp processes and an important question of
current interest is what effects these have both on the ground state
and excitations of the ladders.

A second example in which Umklapp processes may be  important is
$x=1/8$ doped La$_{2-x}$Sr$_x$CuO$_4$\cite{berg_et_al_2007}. 
In this material regular "stripe" order is formed below a critical
temperature ~\cite{Stripes}. Stripes in neighboring  
planes are perpendicular to each other and are shifted by one lattice
spacing~\cite{Tranquada1}. The unit cell in the CuO planes contains
four sites, which can be thought of as forming two undoped and two
doped chains of atoms. Hence the period in the direction perpendicular
to the CuO planes is four. On the other hand, the doped chains are 3/4
filled. As a result the period of the potential induced by the
neighboring planes is also four, which coincides with the average distance
between holes in the doped chains. A simple model describing this
situation is given by doped 2-leg ladders in presence of a periodic
potential. It is well established that La$_{1.875}$Sr$_{0.125}$CuO$_4$
exhibits rather exotic 2D superconducting behavior as a result of the
CuO planes being effectively decoupled from one another
\cite{Tranquada2,Basov}. Similar dynamical layer decoupling  
has recently been observed in  heavy fermion superconductor
CeRhIn$_5$~\cite{CeRhIn}.  

Umklapp processes can in principle also be induced by imposing
external periodic potentials. This has recently been demonstrated by
adsorbing noble gas monolayers on the surface on carbon nanotubes
\cite{cobden}.

From a theoretical point of view, there is one particular case, in
which it is known that Umklapp processes have very interesting physical
consequences. This occurs in the so-called Kondo-Heisenberg model
\cite{zachar,berg}. The latter describes a situation where the 
two legs of the ladder are \emph{inequivalent}. Leg 1 is half-filled
and as a consequence of Umklapp interactions has a large Mott gap,
while leg 2 has a density of less than one electron per site. At low
energies tunneling between the legs is not allowed due to the presence
of a large Mott gap in leg 1, but virtual processes lead to a
Heisenberg exchange interaction between electron spins on the two legs.
The resulting model describing the low-energy physics of such a 2-leg
ladder consists of a spin S=1/2 Heisenberg chain (leg 1) interacting
via exchange interactions with a one-dimensional electron gas (1DEG,
leg 2). Generically the Fermi momentum of the 1DEG will be
incommensurate with the lattice. It was demonstrated in \cite{zachar}
that this Kondo-Heisenberg model exhibits quasi-long-range order of
particular composite order parameter at a finite wave vector. More
recently it was shown \cite{berg} that there also is quasi-long-range
superconducting order with wave number $\pi$, consituting an
example of a 1D Fulde-Ferrell-Larkin-Ovchinnikov state \cite{FFLO} in the
absence of a magnetic field. In very recent work \cite{FradkinArx} it
was demonstrated that the PDW state is in fact much more general and
in particular does not require the legs to be inequivalent. 

\begin{figure}[ht]
\begin{center}
\begin{tabular}{ccc}
(a)\includegraphics[height=0.25\textheight]{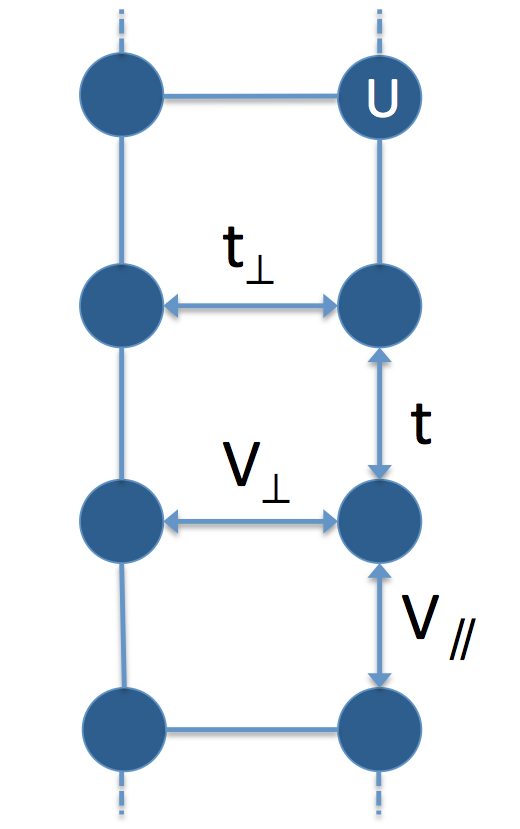}&
\qquad\qquad\qquad&
(b)\includegraphics[width=0.35\textwidth]{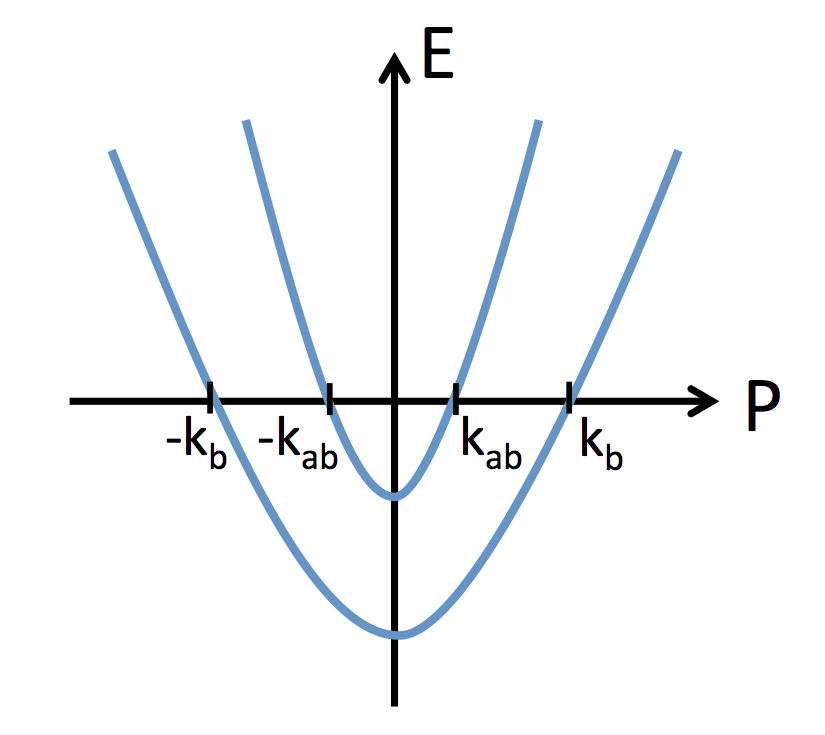}
\end{tabular}
\caption{(a): Extended Hubbard Ladder with different leg and
  rung hopping amplitudes and density-density
  interactions. (b): The non-interacting band structure for
  the tight-binding model on the ladder, with the Fermi wavevectors
  labelled.} 
\label{fig:ladder}
\end{center}
\end{figure}

In the following we consider spin-1/2 fermions on a two-leg ladder
with Hubbard and nearest-neighbor density-density interactions. In
addition we allow an external periodic potential to be present. The
Hamiltonian is given by
\begin{eqnarray}
H_{\rm ladd}(K) &=& -t\sum_{n,\alpha } \sum_{j=1}^2a^\dagger_{j,n+1,\alpha}a_{j,n,\alpha}
+a^\dagger_{j,n,\alpha}a_{j,n+1,\alpha}
-t_\perp\sum_{n,\alpha } a^\dagger_{1,n,\alpha}a_{2,n,\alpha}+
a^\dagger_{2,n,\alpha}a_{1,n,\alpha}+U\sum_{j,l}n_{j,l,\uparrow}n_{j,l,\downarrow}\nn
&+& V_\perp\sum_{l} n_{1,l}n_{2,l} + V_\parallel\sum_{j,l} n_{j,l}n_{j,l+1}+ \sum_{j,l}W_{j} \cos(Kl)\ n_{j,l},
\label{Hamiltonian}
\end{eqnarray}
where $a_{j,n,\alpha}$ are annihilation operators for spin-$\alpha$
electrons on site $n$ of leg $j$ of the ladder and 
$n_{j,l,\alpha}=a^\dagger_{j,l,\alpha}a_{j,l,\alpha}$.
$U$ is the Hubbard interaction strength, $V_\perp$ and $V_{\parallel}$ are
the density-density interaction strengths along the rung and leg
directions respectively and the periodic potential is characterized by its
strength on each leg $W_{1,2}$ and the wavenumber of its modulation,
$K$. The lattice model~(\ref{Hamiltonian}) has U(1)$\times$SU(2)
symmetry, with an additional $\mathbb{Z}_2$ symmetry if $W_1=W_2$. 
It is useful to rewrite the periodic potential term as
\be
\sum_l \cos(Kl)\Big[W_+ (n_{1,l}+n_{2,l})+
W_-(n_{1,l}-n_{2,l})\Big] ,
\label{PP}
\ee
where $W_\pm=(W_1\pm W_2)/2$. A nonzero $W_-$ breaks the symmetry
between the two legs of the ladder. In the following we consider a
case where $W_-=0$ (``$4k_b$ Umklapp'') and one where $W_+=0$
(``$3k_b+k_{ab}$ Umklapp'').
A schematic diagram of the ladder geometry can be seen in Fig.~\ref{fig:ladder}.
In order see which wave numbers $K$ will lead to the most pronounced
effects for weak interactions and small $W_{1,2}$ it is useful to
consider the band structure of $H_{\rm ladd}$ in the absence of
interactions. It is useful to introduce the bonding ($b$) and
antibonding ($ab$) variables by 
\be
c_{j,n,\alpha} =\frac{1}{\sqrt{2}}\left(a_{1,n,\alpha}-(-1)^j
  a_{2,n,\alpha}\right)\ , 
\label{babvar}
\ee
where $j=1,2=b,ab$. In terms of these operators the non-interacting
tight-binding Hamiltonian $H_{\rm ladd,0}$ is diagonal in momentum space  
\be
H_{\rm ladd,0}=\sum_{j=1}^2\sum_k \epsilon_j(k)
c^\dagger_{j,\alpha}(k)c_{j,\alpha}(k)\ ,
\label{diagHam}
\ee
where $c_{j,\alpha}(k)=L^{-1/2}\sum_ne^{ikn}c_{j,n,\alpha}$ and
\be
\epsilon_1(k)=-2t\cos(k)-t_\perp\ ,\quad
\epsilon_2(k)=-2t\cos(k)+t_\perp\ .
\ee
The corresponding band structure is shown in Fig.~\ref{fig:ladder}(b). 
For weak interactions the low energy degrees of freedom occur in the
vicinities of $nk_{b}$ and $nk_{ab}$ where $n$ in an integer and
$k_{b}$, $k_{ab}$ are the Fermi momenta of the bonding and antibonding
bands respectively. It is then clear that external potentials with
wave numbers $K=n_1k_b+n_2k_{ab}$ will affect the low-energy degrees
of freedom most strongly. In the following we concentrate on the cases
$K=3k_b+k_{ab}$ and $K=4k_{b}$. As we will see, in the case of strong
interactions but small $t_\perp$ an analogous picture applies.

This paper is organized as follows. In Section~\ref{sec:Models} we
derive the low-energy effective field theories in the ``band'' and
``chain'' limits of the Hamiltonian~(\ref{Hamiltonian}) and discuss
how we account for the external periodic potential. 
In Section~\ref{sec:4kb} we consider the $4k_b$ Umklapp process in
both band and chain representations of the model. By means of
renormalization group methods we derive the
effective low energy theories describing the strong coupling fixed
points. In Section~\ref{sec:3kb+kab} we analyze the effects of the
$3k_b+k_{ab}$ Umklapp process at low energies in both band and chain 
representations of the model. Section~\ref{sec:DMRG} presents
density matrix renormalization group (DMRG) calculations in
intermediate parameter regimes. Section~\ref{sec:Conclusions} contains
the conclusions. A number of technical points are discussed in several
appendices.

\section{Low-Energy Description}
\label{sec:Models}
There are two complementary ways of deriving a field theory
description of the lattice Hamiltonian~\fr{Hamiltonian}, each 
of which applies to a particular limit of the model. One may start by
considering the non-interacting  Hamiltonian, diagonalizing the
tight-binding model by transforming to bonding and antibonding  
variables and subsequently treating the interaction using
perturbative renormalization group methods
\cite{Fabr93,balents,schulz96,lin}. Hereinafter this approach will be
called the ``band representation''. Alternatively, one may start by
considering two strongly interacting uncoupled chains and treat the
across rung hopping $t_\perp$ and density-density interaction
$V_\perp$ as perturbations \cite{dima,amt}. This approach will be
referred to as the ``chain representation''. In the following
subsections we summarize both approaches in turn.

\subsection{The Band Representation $U, V_j\ll t,t_\perp$}
Here the starting point is the tight-binding model obtained by
dropping all interaction terms in the Hamiltonian \fr{Hamiltonian}.
The resulting model is diagonalized in terms of the bonding and
antibonding ($b/ab$) variables~\fr{babvar}, resulting in split bonding
and antibonding bands \fr{diagHam} as depicted in Fig.~\ref{fig:ladder}(b). 
As we are interested in the low-energy behaviour of the system, we
linearize the spectrum around the Fermi points. The low-energy
projections of the lattice fermion operators are then  
\be
c_{j,n,\alpha}\sim\sqrt{a_0}\left[R_{j,\alpha}(x)e^{ik_{j}x}+L_{j,\alpha}(x)e^{-ik_{j}x}\right],
\label{continuum}
\ee
where $L(x)$ and $R(x)$ are left and right moving fermion fields close
to the Fermi points,  
$k_b$ ($k_{ab}$) is the Fermi wavevector in the bonding (antibonding) band and
$a_0$ is the lattice spacing, which serves as the short-distance
cut-off of the theory. The interactions are conveniently expressed in
terms of currents \cite{balents}, which following Ref.~\cite{CCL}  we define as
\bea
I_{ij}^R&=&\frac{1}{2}R_{i,\alpha}\ \epsilon_{\alpha\beta}\ R_{j,\beta}\
,\qquad
I^{aR}_{ij}=\frac{1}{2}R_{i,\alpha}(\epsilon\sigma^a)_{\alpha\beta}
R_{j,\beta}\ ,\\
J_{ij}^R&=&\frac{1}{2}R^\dagger_{i,\alpha}\  R_{j,\alpha}\ ,\qquad~~\quad
J^{aR}_{ij}=\frac{1}{2}R^\dagger_{i,\alpha}\sigma^a_{\alpha\beta}  R_{j,\beta},
\label{eqs:bilinears}
\eea
and similarly for left-moving fermion fields with $R\leftrightarrow L$.
The low-energy Hamiltonian then takes the form
$H=\int dx\ \left[{\cal H}_0+{\cal H}_U+{\cal H}_W\right]$, where
\bea
{\cal H}_0&=&\sum_{j=1}^2v_j\Big(-iR^\dagger_{j,\alpha}\partial_x
R_{j,\alpha}
+iL^\dagger_{j,\alpha}\partial_x L_{j,\alpha}\Big)\ ,\nn
{\cal H}_U&=&\sum_{i,j}\tilde{c}^\rho_{ij}J_{ij}^RJ_{ij}^L
-\tilde{c}^\sigma_{ij}J^{aR}_{ij}J^{aL}_{ij}
+\sum_{i\neq j}\tilde{f}^\rho_{ij}J_{ii}^RJ_{jj}^L
-\tilde{f}^\sigma_{ij}J^{aR}_{ii}J^{aL}_{jj}\ ,\nn
{\cal H}_W&=&\sum_{P\in S}\sum_{\sigma=\pm}W_{\sigma,P}\
\delta_{K,P}\left[\rho^{(\sigma)}_{P}(x)+ 
{\rm h.c.}\right].
\label{eqs:FieldHamiltonian}
\eea
Here $\rho^{(\sigma)}_{P}(x)$ are the Fourier components of the
low-energy projections of $n_{1,l}\pm n_{2,l}$, c.f. Eq.~\fr{PP}, 
with momenta close to $P$;
these components are discussed in some detail in
Appendix~\ref{app:CD}. In the following 
we consider ``$4k_F$'' components with wave numbers around
$S=\{4k_{b},~4k_{ab},~3k_{b}+k_{ab},~3k_{ab}+k_b\}$. 
The ``$2k_F$''-response is generally blocked by the presence of a spin gap
in doped Hubbard ladders, see e.g. Appendix~\ref{app:2kF}, and
we shall not consider them here. The $4k_F$
components of the density are obtained by integrating out the
high-energy degrees of freedom perturbatively in $U$, see
Appendix~\ref{app:4kF}, and are given in terms of the currents as
\bea
\rho^{(+)}_{4k_b}(x)+{\rm h.c.}&=& \big(I_{11}^L\big)^\dagger
I_{11}^R+\big(I_{11}^R\big)^\dagger I_{11}^L\ ,\nn
\rho^{(+)}_{4k_{ab}}(x)+{\rm h.c.}
&=&\big(I_{22}^L\big)^\dagger I_{22}^R+\big(I_{22}^R\big)^\dagger I_{22}^L\ ,\nn
\rho^{(+)}_{2k_b+2k_{ab}}(x)+{\rm h.c.}&=&
8\left\{\big(I_{12}^L\big)^\dagger I_{21}^R+\big(I_{21}^R\big)^\dagger
  I_{12}^L\right\} ,\nn 
\rho^{(-)}_{k_b+3k_{ab}}(x)+{\rm h.c.}
&=&2\left\{\big(I_{22}^L\big)^\dagger I^R_{21}+
\big(I_{21}^L\big)^\dagger I^R_{22}+
\big(I_{22}^R\big)^\dagger I^L_{21}+
\big(I_{21}^R\big)^\dagger I^L_{22}\right\}\ ,\nn
\rho^{(-)}_{3k_b+k_{ab}}(x)+{\rm h.c.}&=&
2\left\{\big(I_{11}^L\big)^\dagger I^R_{12}+
\big(I_{12}^L\big)^\dagger I^R_{11}+
\big(I_{11}^R\big)^\dagger I^L_{12}+
\big(I_{12}^R\big)^\dagger I^L_{11}\right\}.
\eea
The initial conditions for the coupling constants defined in
\fr{eqs:FieldHamiltonian} for the extended Hubbard model are
\bea
\tilde{c}_{ii}^\rho &=& U+V_\perp +4V_\parallel\left[1-\frac{1}{2}\cos(2k_{i}a_0)\right],\nn
\tilde{c}_{ij}^\rho &=& U-V_\perp +4V_\parallel\left[\cos((k_{b}-k_{ab})a_0)-\frac{1}{2}\cos((k_{b}+k_{ab})a_0)\right],\nn
\tilde{f}_{ij}^\rho &=& U+3V_\perp +4V_\parallel\left[1-\frac{1}{2}\cos((k_b+k_{ab})a_0)\right],\nn
\tilde{c}_{ii}^\sigma &=& U+V_\perp +2V_\parallel\cos(2k_ia_0),\nn
\tilde{f}_{ij}^\sigma &=& \tilde{c}_{ij}^\sigma = U-V_\perp +2V_\parallel\cos((k_b+k_{ab})a_0).\nonumber
\label{eqs:ExtHubInitConds}
\eea

The analysis which we carry out in the band representation requires
the bosonized Hamiltonian. Following Ref.~\cite{GNT}, we bosonize the
Hamiltonian according to
\be
R_{d,\sigma}\sim \frac{\kappa_{d,\sigma}}{\sqrt{2\pi}}
e^{i\sqrt{4\pi}\varphi_{d,\sigma}}\ ,\qquad
L_{d,\sigma}\sim\frac{\kappa_{d,\sigma}}{\sqrt{2\pi}}
e^{-i\sqrt{4\pi}\bar{\varphi}_{d,\sigma}},\qquad d=1,2=b,ab,
\label{bosonization1}
\ee
where $\varphi_{d,\sigma}$ ($\bar\varphi_{d,\sigma}$) is the right (left) chiral component of
a canonical boson field and $\{\kappa_{d,\sigma},\kappa_{d',\sigma'}\}=2\delta_{d,d'}\delta_{\s,\s'}$ are
Klein factors to ensure the anti-commutation of different species of fermions.  The boson
fields have commutation relations
\be
[\varphi_{d,\s}(x),\varphi_{d',\s'}(x')]=-[\bar\varphi_{d,\s}(x),\bar\varphi_{d',\s'}(x')]=\frac{i}{4}{\rm sgn}(x-x')\delta_{d,d'}\delta_{\s,\s'},\quad
[\varphi_{d,\s}(x),\bar\varphi_{d',\s'}(x')]=\frac{i}{4}\delta_{d,d'}\delta_{\s,\s'},
\ee
which enforce anti-commutation relations for fermions of the same species.
 Then, we change to spin and charge bosons according to
\bea
\Phi_{d,c} &=& \frac{1}{\sqrt{2}}\left[\varphi_{d,\uparrow}+\varphi_{d,\downarrow}+ \bar\varphi_{d,\uparrow}+\bar\varphi_{d,\downarrow}\right],\quad
\Phi_{d,s} = \frac{1}{\sqrt{2}}\left[\varphi_{d,\uparrow}-\varphi_{d,\downarrow}+ \bar\varphi_{d,\uparrow}-\bar\varphi_{d,\downarrow}\right],\nn
\Theta_{d,c}&=& \frac{1}{\sqrt{2}}\left[\varphi_{d,\uparrow}+\varphi_{d,\downarrow}- \bar\varphi_{d,\uparrow}-\bar\varphi_{d,\downarrow}\right],\quad
\Theta_{d,s} = \frac{1}{\sqrt{2}}\left[\varphi_{d,\uparrow}-\varphi_{d,\downarrow}- \bar\varphi_{d,\uparrow}+\bar\varphi_{d,\downarrow}\right],
\label{spinchargefields}
\eea
where $\Phi$ and $\Theta$ are dual bosons obeying $[\Theta(x),\Phi(x')]=i\vartheta(x-x')$, where $\vartheta(y)$ 
the Heaviside step function. This relationship also implies that $[\partial_x\Theta(x),\Phi(x')]=i\delta(x-x')$ are canonically
conjugate. 
The resulting bosonized Hamiltonian is given by
\bea
{\cal H}_0+{\cal H}_U&=&\sum_{d=1}^2\frac{v_d}{2\pi}
\sum_{\gamma=c,s}
\left[\left(\partial_x\Phi_{d,\gamma}\right)^2+
\left(\partial_x\Theta_{d,\gamma}\right)^2\right]
+\frac{\tilde{c}_{dd}^\rho}{(2\pi a_0)^2}\frac{1}{8\pi}
\left[\left(\partial_x\Phi_{d,c}\right)^2-
\left(\partial_x\Theta_{d,c}\right)^2\right]\nn
&+&\frac{\tilde{f}_{12}^\rho}{4\pi(2\pi a_0)^2}
\left[\partial_x\Phi_{1,c}\partial_x\Phi_{2,c}-
\partial_x\Theta_{1,c}\partial_x\Theta_{2,c}\right]
+\sum_{d=1}^2
\frac{\tilde{c}_{dd}^\sigma}{(2\pi a_0)^2}\left[
\cos\big(\sqrt{8\pi}\Phi_{d,s}\big)-
\frac{1}{8\pi}
\left[\left(\partial_x\Phi_{d,s}\right)^2-
\left(\partial_x\Theta_{d,s}\right)^2\right]\right]\nn
&+&\frac{2\tilde{f}^\sigma_{12}}{(2\pi a_0)^2}\left[\cos\Big(\sqrt{2\pi}
\big(\Phi_{1,s}+\Phi_{2,s}\big)\Big)\
\cos\Big(\sqrt{2\pi}
\big(\Theta_{1,s}-\Theta_{2,s}\big)\Big)-\frac{1}{8\pi}\Big(\partial_x\Phi_{1,s}\partial_x\Phi_{2,s}-\partial_x\Theta_{1,s}\partial_x\Theta_{2,s}\Big)\right]\nn
&+&\left[
\frac{\tilde{c}^\rho_{12}-\tilde{c}^\sigma_{12}}{(2\pi a_0)^2}
\cos\Big(\sqrt{2\pi}\big(\Theta_{1,s}-\Theta_{2,s}\big)\Big)
-\frac{\tilde{c}^\rho_{12}+\tilde{c}^\sigma_{12}}{(2\pi a_0)^2}
\cos\Big(\sqrt{2\pi}
\big(\Phi_{1,s}-\Phi_{2,s}\big)\Big)\right]
\cos\Big(\sqrt{2\pi}
\big(\Theta_{1,c}-\Theta_{2,c}\big)\Big)
\nn
&-&\frac{2\tilde{c}^\sigma_{12}}{(2\pi a_0)^2}
\cos\Big(\sqrt{2\pi}
\big(\Theta_{1,c}-\Theta_{2,c}\big)\Big)
\cos\Big(\sqrt{2\pi}
\big(\Phi_{1,s}+\Phi_{2,s}\big)\Big).
\label{eq:BosonHamiltonian}
\eea

There is a convenient way to classify the ground state phase of the
ladder in terms of the spin and charge bosons. Following
Ref.~\cite{balents}, phases will be classified by the number of spin
and charge bosons which remain gapless. In particular, we will use the
notation $CmSn$ where $m$ is the number of gapless charge bosons and
$n$ is the number of gapless spin bosons.

\subsection{The Chain Representation $t_\perp\ll t,U,t^2/U$}
\label{sec:chain}
The field theory for the chain representation of \fr{Hamiltonian} is
derived in a succession of steps, outlined below; a detailed
derivation can be found in Ref.~\cite{amt}. An important feature of
the chain representation is that longer range density-density
interactions along the chain direction
\be
\sum_{j,l} \sum_{m\geq 2} V_{\parallel,m} n_{j,l}n_{j,l+m}
\ee
can be easily accommodated. As long as $V_{\parallel,m}$ are
sufficiently small and decreasing with $m$, the main effect of this
extended interaction is to decrease the value of $K_c$ in \fr{FT}.
We will make use of this device for tuning the value of $K_c$ in the
following. 

The main assumption of the derivation is that the interchain hopping
$t_\perp$ is small in comparison to the high-energy cutoffs, which for
$V_\perp,V_\parallel\ll U$ are given by the single chain band-width
and the exchange energy scale ($\sim t^2/U$ at large $U$). The
Hamiltonian is first bosonized for $t_\perp=V_{\perp}=W_{1,2}=0$ using
standard results for the one-dimensional (extended) Hubbard model
\cite{book,GNT}. The resulting theory (as long as $V_\parallel$ is
not too large) is the sum of four Gaussian models for
spin and charge bosonic fields in each chain. Denoting the bosonic
fields by $\Phi^{(i)}_j$ where $i=c,s$ denotes the spin or charge
sector and $j=1,2$ denotes the chain, we form symmetric and
antisymmetric combinations of the fields
 \bea
 && \Phi_c = \frac{1}{\sqrt{2}}(\Phi^{(c)}_1 + \Phi^{(c)}_2), ~~ \Phi_f =
 \frac{1}{\sqrt{2}}(\Phi^{(c)}_1 - \Phi^{(c)}_2),\nonumber\\ 
 && \Phi_s = \frac{1}{\sqrt{2}}(\Phi^{(s)}_1 + \Phi^{(s)}_2), ~~ \Phi_{sf }=
\frac{1}{\sqrt{2}}(\Phi^{(s)}_1 - \Phi^{(s)}_2).  
\label{fields_chain}
 \eea
In the absence of a periodic potential and away from commensurate
fillings, the $\Phi_c$ field decouples from the other fields. It is
then described by a Gaussian (Tomanaga-Luttinger) theory with the
Hamiltonian density 
\be
{\cal H}_c = \frac{v_c}{2}\Big[K_c(\p_x\Theta_c)^2
+K_c^{-1}(\p_x\Phi_c)^2\Big],  
\label{Hc}
\ee
where $K_c<1$ is the Luttinger parameter in the charge sector and $v_c$
is the charge velocity. The exact dependence of these parameters on
the underlying lattice parameters is complicated, but for
$V_\parallel=0$ can be extracted from the exact solution of the
one-dimensional Hubbard model~\cite{book,chubukov}. 

The remaining bosonic fields are refermionized in terms of six
Majorana fermion fields. For the right-moving components we have
\bea
&& \chi_R^0 = \frac{ \kappa_{sf}}{\sqrt{\pi a_0}}\sin(\sqrt{4\pi}\phi_{sf}), ~~ \chi_R^3 = \frac{ \kappa_{sf}}{\sqrt{\pi a_0}}\cos(\sqrt{4\pi}\phi_{sf}),\nn
 && \chi_R^1 = \frac{ \kappa_{s}}{\sqrt{\pi
     a_0}}\sin(\sqrt{4\pi}\phi_s), ~~ \chi_R^2 = \frac{
   \kappa_{s}}{\sqrt{\pi a_0}}\cos(\sqrt{4\pi}\phi_s)\ ,\nonumber\\ 
 && \xi_R^3 = \frac{ \kappa_{c}}{\sqrt{\pi
     a_0}}\sin(\sqrt{4\pi}\phi_f), ~~ \eta_R = \frac{
   \kappa_{c}}{\sqrt{\pi a_0}}\cos(\sqrt{4\pi}\phi_f)\ ,
 \eea
where $\phi_a$ are the right-moving chiral components of the canonical Bose fields
$\Phi_a$ ($a=f,s,sf$) and $\kappa_{a}$ are Klein factors fulfilling 
$\{\kappa_a,\kappa_b\}=2\delta_{a,b}$. Analogous expressions with $R$ replaced by 
$L$ and $\phi$ by $\bar\phi$ hold for the left-moving modes.

The next step of the derivation introduces the interchain tunneling $t_\perp$. This induces
a hybridization between the $\eta$ and $\chi^0$ fermions. Following \cite{amt} we examine the
part of the Hamiltonian which is quadratic in terms of the $\eta$ and $\chi^0$ Majorana fermions.
We linearize the spectrum about the wavevector $Q=t_\perp/\sqrt{v_cv_s}$ where $E(Q)=0$ and introduce 
the new Majorana fermions $\xi_{R,L}^{1,2}$ which diagonalize the aforementioned quadratic 
part of the Hamiltonian. The new Majorana fermions are given by
\bea
\begin{pmatrix}
\chi_R^0\cr
\eta_R
\end{pmatrix}&=&
\sqrt{\frac{2}{v_s+v_c}}
\begin{pmatrix}
\sqrt{v_c}\cos(Qx) & \sqrt{v_c}\sin(Qx)\cr
-\sqrt{v_s}\sin(Qx) & \sqrt{v_s}\cos(Qx)
\end{pmatrix}
\begin{pmatrix}
\xi_R^1\cr
\xi_R^2
\end{pmatrix}\ ,
\label{MajoanatrafoR}\nn
\begin{pmatrix}
\chi_L^0\cr
\eta_L
\end{pmatrix}&=&
\sqrt{\frac{2}{v_s+v_c}}
\begin{pmatrix}
\sqrt{v_c}\cos(Qx) & -\sqrt{v_c}\sin(Qx)\cr
\sqrt{v_s}\sin(Qx) & \sqrt{v_s}\cos(Qx)
\end{pmatrix}
\begin{pmatrix}
\xi_L^1\cr
\xi_L^2
\end{pmatrix}.
\label{MajoanatrafoL}
\eea
In terms of these new variables the low-energy Hamiltonian takes the
form ${\cal H} = {\cal H}_c+{\cal H}_0 + {\cal V}_{\rm int}+{\cal H}_W$,
where
\bea
\label{FT}
{\cal H}_c &=& \frac{v_c}{2}\Big[K_c(\p_x\Theta_c)^2 +K_c^{-1}(\p_x\Phi_c)^2\Big], \label{H_c}\\ 
{\cal H}_0 &=& \frac{\ri v_c}{2}(\xi^3_L \p_x \xi^3_L-
\xi_R^3\p_x\xi^3_R) 
+\frac{\ri u}{2}\sum_{a=1,2}(\xi^a_L\p_x \xi_L^a - \xi_R^a\p_x
  \xi_R^a)  +  \frac{\ri v_s}{2}\sum_{a=1}^3(\chi^a_L\p_x \chi_L^a -
  \chi_R^a\p_x \chi_R^a)\ ,\label{H0}\\
{\cal V}_{\rm int} &=&
-2(\xi^3_R\xi^3_L)\Big[g_{\s,-}(\chi_R^a\chi_L^a) +
g_{c,ss}(\xi_R^1\xi_L^1 - \xi_R^2\xi_L^2)\Big] -
g_{\rho,-}(\xi^1_R\xi_L^1 - \xi_R^2\xi_L^2)^2 \nonumber\\ 
&& - 2g_{c,st}(\xi^1_R\xi_L^1 - \xi_R^2\xi_L^2)\sum_{a=1}^3(\chi_R^a\chi_L^a) - 2g_{\s,+}\sum_{a>b,a,b=1}^3(\chi_R^a\chi_L^a)(
 \chi_R^b\chi_L^b) \label{Int}\ ,\nn
{\cal H}_W&=&\sum_{P\in S}\sum_{\sigma=\pm}W_{\sigma,P}\
\delta_{K,P}\left[\rho^{(\sigma)}_{P}(x)+ 
{\rm h.c.}\right].
 \eea
Here $v_{c,s}$ are the charge and spin velocities of uncoupled
chains, $S=\{4k_F,4k_F\pm Q,4k_F\pm 2Q\}$ and
\be
u=\frac{2v_cv_s}{v_s+v_c}.
\ee
The Hamiltonian ${\cal H}_c+{\cal  H}_0+{\cal V}_{\rm int}$  has the
same symmetry U(1)$\times$SU(2)$\times\mathbb{Z}_2$ as the underlying
lattice model for $W_{1,2}=0$.
The coupling parameters of the continuum Hamiltonian are determined by the 
underlying lattice model~\fr{Hamiltonian}
\bea
g_{\sigma -} &=& \frac{\alpha V_\perp}{2},\qquad
g_{\sigma +} = \frac{1}{2} \pi v_s g_{\lambda},\qquad
g_{c,ss} = u \left(\frac{\alpha}{v_s} V_\perp - 2g_k\right),\nn
g_{c,st} &=& u\left(\frac{\alpha}{v_c} V_\perp + \pi g_\lambda\right),\qquad
g_{\rho,-} = \frac{v_sv_c}{(v_s+v_c)^2}\alpha V_\perp,
\eea
where $\alpha$ is a short-distance cut-off, $g_k$ characterizes the four-fermion
interaction in the $\Phi_f$ sector, which for $|K_c-1|\ll1$ is given
by $g_k\approx2\pi\left(1/K_c-1\right)$, and $g_{\lambda}$ is the
strength of the marginally irrelevant spin-current interaction for a
single extended Hubbard chain, which is known only for
small $U$ and $V_\parallel$. The notable differences between this
formulation and the band representation is the presence of several
different velocities $v_c\neq v_s\neq u$; for large intrachain
interactions these differences can be significant. The low-energy
projections of the periodic potential with wave numbers close to
$4k_F$ are derived in Appendix \ref{app:4kFchain}
\bea
\rho^{(+)}_{4k_{F}}(x)&\sim& \frac{\ri F}{2}e^{i\sqrt{4\pi}\Phi_c}
\Big\{\xi^3_R\xi^3_L + \frac{v_s}{(v_s +
  v_c)}[\xi^1_{R}\xi^1_{L} - \xi^2_{R}\xi^2_{L}]\Big\}\ , \label{4kF}\\ 
\rho^{(+)}_{4k_{F}-2Q}(x)&\sim&
\frac{iv_s F}{2(v_s + v_c)}e^{i\sqrt{4\pi}\Phi_c}\ (\xi_L^1
-\ri\xi_L^2)(\xi_R^1 +\ri\xi_R^2)\ ,\label{4-2Q} \\
\rho^{(+)}_{4k_{F}+2Q}(x)&\sim&
\frac{iv_s F}{2(v_s + v_c)}e^{i\sqrt{4\pi}\Phi_c}\ (\xi_R^1
-\ri\xi_R^2)(\xi_L^1 +\ri\xi_L^2)\ ,\label{4+2Q} \\
\rho^{(-)}_{4k_F-Q}(x) &=& -iF  \sqrt{\frac{v_s}{2\left(v_s+v_c\right)}}e^{i\sqrt{4\pi}\Phi_{c}}\left[\left(\xi_R^1+i\xi_R^2\right)\xi_L^3+\xi_R^3 \left(\xi_L^1-i\xi_L^2\right)\right],\\
\rho^{(-)}_{4k_F+Q}(x) &=& -iF \sqrt{\frac{v_s}{2\left(v_s+v_c\right)}} e^{i\sqrt{4\pi}\Phi_{c}}\left[\left(\xi_R^1-i\xi_R^2\right)\xi_L^3+\xi_R^3\left(\xi_L^1+i\xi_L^2\right)\right].
\label{4+Q}
\eea
We note that $\rho^{(+)}_{4k_F}(x)$ and $\rho^{(+)}_{4k_F\pm 2Q}(x)$ are even
under interchange of chains 1 and 2, while $\rho^{(-)}_{4k_F\pm Q}(x)$ are
odd. 
\subsection{Correspondence between chain and band representations}
The correspondence between chain and band representations is as follows
\bea
4k_{F}&\leftrightarrow& 2(k_b+k_{ab})\ ,\nn
4k_{F}+2Q&\leftrightarrow& 4k_b\ ,\nn
4k_{F}-2Q&\leftrightarrow& 4k_{ab}\ ,\nn
4k_{F}+Q&\leftrightarrow& 3k_b+k_{ab}\ ,\nn
4k_{F}-Q&\leftrightarrow& 3k_{ab}+k_b\ .
\label{eqs:dictionary}
\eea
Without lose of generality, we will consider the $4k_{F}+Q$ and
$4k_{F}+2Q$ Umklapp scattering processes. The following analyses are
easily performed for $Q\rightarrow-Q$ and yield analogous results.

\section{$4k_b$ Umklapp}
\label{sec:4kb}
In this section we consider the $4k_b$ Umklapp scattering
process. This may become activated at commensurate filling within the
bonding band \cite{balents} or at incommensurate fillings for an
applied external potential modulated at $4k_b$. In the following we
analyze band and chain limits of \fr{Hamiltonian} in turn and
discuss the zero temperature phase diagram.
The $4k_b$ Mott insulating phase in the two-leg ladder has been
analyzed using RG in the band representation in a very recent work by
Jaefari and Fradkin~\cite{FradkinArx}, which appeared while our
manuscript was being completed. The main result of this analysis is
the existence of a pair-density wave phase. As our discussion differs
substantially (both in details of the RG procedure, the derivation of
the low-energy projections of observables and the analysis of dominant
correlations), we nevertheless present it in detail in the following. 
\subsection{Band Representation}
Here our general approach is to consider the 1-loop renormalization
group (RG) equations for the Hamiltonian~\fr{eqs:FieldHamiltonian}
in presence of the $4k_b$ Umklapp interaction term. In the field
theory limit the latter becomes
\be
{\cal H}_W=\frac{\tilde{u}^\rho_{11}}{2}\int dx\ 
\left[(I^L_{11})^\dagger I^R_{11}+(I^R_{11})^\dagger
I^L_{11}\right].
\ee
In the notations of Refs.~\cite{lin,CCL}, the one-loop RG equations
are
\bea
\dot{c}^\rho_{11}&=&
-\frac{\alpha}{4}\left[
(c^\rho_{12})^2+3(c^\sigma_{12})^2\right]+\big(u^\rho_{11}\big)^2\ ,\nn
\dot{c}^\rho_{12}&=&-\frac{1}{4}\left[
c^\rho_{11}c^\rho_{12}+3c^\sigma_{11}c^\sigma_{12}\right]
-\frac{1}{4}\left[
c^\rho_{12}c^\rho_{22}+3c^\sigma_{12}c^\sigma_{22}\right]
+\frac{1}{2}\left[c^\rho_{12}f^\rho_{12}+3c^\sigma_{12}f^\sigma_{12}\right]
\ ,\nn
\dot{c}^\rho_{22}&=&-\frac{\alpha}{4}\left[
(c^\rho_{21})^2+3(c^\sigma_{21})^2\right]\ ,\nn
\dot{c}^\sigma_{11}&=&-(c^\sigma_{11})^2
-\frac{\alpha}{2}c^\sigma_{12}(c^\rho_{12}+c^\sigma_{12})\ , \nn
\dot{c}^\sigma_{12}&=&-\frac{1}{4}\left[(c^\rho_{11}+
c^\rho_{22})c^\sigma_{12}+(c^\rho_{12}+2c^\sigma_{12})
(c^\sigma_{11}+c^\sigma_{22})\right]
+\frac{1}{2}\left[c^\rho_{12}f^\sigma_{12}+c^\sigma_{12}f^\rho_{12}
-2c^\sigma_{12}f^\sigma_{12}\right]\ ,\nn
\dot{c}^\sigma_{22}&=&-(c^\sigma_{22})^2
-\frac{\alpha}{2}c^\sigma_{12}(c^\rho_{12}+c^\sigma_{12})\ ,\nn
\dot{f}^\rho_{12}&=&\frac{1}{4}\left[
(c^\rho_{12})^2+3(c^\sigma_{12})^2\right]\ ,\nn
\dot{f}^\sigma_{12}&=&-(f_{12}^\sigma)^2+\frac{1}{2}
c^\sigma_{12}(c_{12}^\rho-c_{12}^\sigma)\ ,\nn
\dot{u}_{11}^\rho&=&c^\rho_{11}\ u^\rho_{11},
\label{4kbRG}
\eea
where $\alpha= (v_1+v_2)^2/(4v_1v_2)$ and the coupling constants have
been rescaled by $\tilde{g}_{ij}=g_{ij}\pi(v_1+v_2)$. Equations
\fr{4kbRG} agree with the RG equations reported in Ref. \cite{balents}
up to a factor of 2 in the equation for $u_{11}^\rho$.

Further progress is made by numerically integrating these
equations. We consider the case where the Umklapp interaction emerges
at a particular doping of an extended Hubbard ladder. We further
restrict our discussion to (sufficiently) small values of $V_\perp/U$ and
$V_\parallel/U$. Then, the numerical integration of Eqs.~(\ref{4kbRG})
gives 
\be c_{11}^\rho,u_{11}^\rho\rightarrow\infty,\quad
{\rm with }\quad c_{11}^\rho/u_{11}^\rho\rightarrow1,
\ee
whilst all other couplings remain small (their ratios to $c_{11}^\rho$
vanish).

The coupling constants which flow to strong coupling are only in the bonding
charge ($1c$) sector of the bosonized Hamiltonian~\fr{eq:BosonHamiltonian} and 
cause the $\Phi_{1c}$ boson to become massive. Now, we employ two-cutoff
scaling~\cite{GNT}, where we integrate out the now massive $\Phi_{1c}$ boson
and its disordered dual $\Theta_{1c}$ perturbatively in the remaining small
couplings. Expanding the partition function to second order in the small couplings,
we obtain an effective action
\be
S_{\rm eff}\approx \tilde{S}_0+\langle \tilde{S}_{\rm
  int}\rangle_{1,c}
-\frac{1}{2}\left[\langle \tilde{S}^2_{\rm int}\rangle_{1,c}
-\langle \tilde{S}_{\rm int}\rangle_{1,c}^2\right]+\ldots,
\ee
with 
\bea
\langle {\cal O}\rangle_{1,c} &=&
\int {\cal D}\Phi_{1,c}\ e^{-S_{1,c}}\ {\cal O},\\
S_{1c}&=&\int dx d\tau\left\{\big(1+\frac{{c}_{11}^\rho}{8\pi v_1(2\pi
    a_0)^2}\big)
\left[v_1\big(\partial_x\Phi_{1,c}\big)^2+
\frac{1}{v_1}\big(\partial_\tau\Phi_{1,c}\big)^2\right]
-\frac{u_{11}^\rho}{(2\pi a_0)^2}\cos\sqrt{8\pi}\Phi_{1,c}
\right\},\\
\tilde{S}_{\rm int}&=&\int dx d\tau\Bigg\{
\frac{f_{12}^\rho}{(2\pi a_0)^2}\frac{1}{4\pi}
\left[\partial_x\Phi_{1,c}\partial_x\Phi_{2,c}+\frac{1}{v_1v_2}
\partial_\tau\Phi_{1,c}\partial_\tau\Phi_{2,c}\right]\nn
&&\qquad\qquad+
\frac{c^\rho_{12}-c^\sigma_{12}}{(2\pi a_0)^2}
\cos\Big(\sqrt{2\pi}
\big(\Theta_{1,c}-\Theta_{2,c}\big)\Big)\
\cos\Big(\sqrt{2\pi}\big(\Theta_{1,s}-\Theta_{2,s}\big)\Big)\nn
&&\qquad\qquad-\frac{c^\rho_{12}+c^\sigma_{12}}{(2\pi a_0)^2}
\cos\Big(\sqrt{2\pi}
\big(\Theta_{1,c}-\Theta_{2,c}\big)\Big)\
\cos\Big(\sqrt{2\pi}
\big(\Phi_{1,s}-\Phi_{2,s}\big)\Big)\nn
&&\qquad\qquad-2\frac{c^\sigma_{12}}{(2\pi a_0)^2}
\cos\Big(\sqrt{2\pi}
\big(\Theta_{1,c}-\Theta_{2,c}\big)\Big)\ 
\cos\Big(\sqrt{2\pi}
\big(\Phi_{1,s}+\Phi_{2,s}\big)\Big)\Bigg\},
\eea
and $\tilde{S}_0$ describes all other terms in the action which do not feature
$1c$ bosons. The action for the bonding charge boson $S_{1c}$ is an effective
Sine-Gordon model~\cite{GNT}. 
The RG flow of the coupling $u_{11}^\rho$ pins the charge boson
$\Phi_{1c}$ to zero. Thus $\langle\Phi_{1c}\rangle_{1c}=0$ and
two-point functions obey 
\bea
\big\langle e^{i\beta\Theta_{1,c}(\tau,x)}e^{-i\beta'\Theta_{1,c}(\tau',x')}\big\rangle_{1,c} &\propto& \delta_{\beta,\beta'}e^{-r_1/\xi},\nn
\left\langle \partial_y\Phi_{1,c}(\tau,x)\partial_{y'}\Phi_{1,c}(\tau',x')\right\rangle_{1,c} &\propto& \partial_y\partial_{y'}\frac{e^{-2r_1/\xi}}{\left(2r_1/\xi\right)^2},\nn
\big\langle\partial_y\Phi_{1,c}(\tau,x)e^{-i\beta'\Theta_{1,c}(\tau',x')}\big\rangle_{1,c} &=& 0,
\eea
where $y=x,v_1\tau$ and $r_1^2=v_1^2(\tau-\tau')^2+(x-x')^2$. The
first relation follows from topological charge conservation in the
sine-Gordon model and the second follows from the properties of
massive bosons in one-dimensional systems. For all other operator
product expansions we use those of the corresponding Gaussian
models. To second order in the perturbative expansion the effective
Hamiltonian density is of the form 
\bea
{\cal H}_{\rm eff}&=&\frac{\bar{v}_2}{2\pi}
\left[\frac{1}{K_{2,c}}
\left(\partial_x\Phi_{2,c}\right)^2+
K_{2,c}
\left(\partial_x\Theta_{2,c}\right)^2\right]
\nn
&+&
\sum_{d=1}^2\frac{\bar{v}_d}{2\pi}
\left[\left(\partial_x\Phi_{d,s}\right)^2+
\left(\partial_x\Theta_{d,s}\right)^2\right]
+\bar{c}_{dd}^\sigma\left[\cos\big(\sqrt{8\pi}\Phi_{d,s}\big)
-\frac{1}{8\pi}
\left[\left(\partial_x\Phi_{d,s}\right)^2-
\left(\partial_x\Theta_{d,s}\right)^2\right]\right]\nn
&+&2\bar{f}_{12}^\sigma\left[\cos\big(\sqrt{2\pi}(\Phi_{1,s}+\Phi_{2,s})\big)
\ \cos\big(\sqrt{2\pi}(\Theta_{1,s}-\Theta_{2,s})\big)-\frac{1}{8\pi}\Big(
\partial_x\Phi_{1,s}\partial_x\Phi_{2,s}-\partial_x\Theta_{1,s}\partial_x\Theta_{2,s}\Big)\right]\nn
&+&\lambda\cos\big(\sqrt{2\pi}(\Phi_{1,s}-\Phi_{2,s})\big)
\ \cos\big(\sqrt{2\pi}(\Theta_{1,s}-\Theta_{2,s})\big),
\label{4kbEffH}
\eea
where $\lambda$ is a coupling constant generated in the
renormalization group procedure, which is second order in the
remaining small couplings. The $\lambda$-term carries conformal spin
and as a result only has minor effects at weak coupling \cite{zachar}.  
The structure of the low-energy effective field theory $H_{\rm eff}$
is the same as for the Kondo-Heisenberg model~\cite{zachar}. We
therefore can take over the RG analysis of \cite{RGKH} in order to
infer the phase diagram. In the Kondo-Heisenberg model there are two
distinct phases: for ferromagnetic Heisenberg exchange interactions
between the spin-chain and the one-dimensional electron gas (1DEG) the
RG flow is towards weak coupling and approaches a $C1S2$ fixed point,
described by a 3-component Luttinger liquid Hamiltonian for the
$\Phi_{2,c}$, $\Phi_{1,s}$ and $\Phi_{2,s}$ bosons. On the other hand,
for antiferromagnetic Heisenberg exchange interactions between the
spin-chain and the one-dimensional electron gas (1DEG) the RG flow is
towards strong coupling. Spin gaps open in both spin sectors and one
ends up with a $C1S0$ phase. 

Which phase the Hamiltonian \fr{4kbEffH} flows to under RG depends on
the values of the bare couplings and concomitantly the ratios
$V_\parallel/U$ and $V_\perp/U$.

\subsubsection{$C1S2$ Phase}
\label{C1S2}
For Hubbard model initial conditions the RG flow of \fr{4kbEffH} is
always towards weak coupling as discussed by Balents and Fisher
\cite{balents}. This corresponds to ferromagnetic exchange between
the spin-chain and the one-dimensional electron gas (1DEG) in the
Kondo-Heisenberg model. More generally, we find that this phase
occurs for $\check{{f}}_{12}^\s>0$, where $\check{{f}}_{12}^\s$
is the initial value of the coupling ${\bar{f}}_{12}^\s$ after
integrating out the $c,1$ boson in our two-cutoff RG scheme. 
Integrating the RG equations \fr{4kbRG} with extended Hubbard model
initial conditions \fr{eqs:ExtHubInitConds} we observe that the values
of $f_{12}^\s$ after the initial flow in our two-cutoff scheme are
positive, as long as $V_\parallel/U$ and $V_\perp/U$ are sufficiently
small. Assuming that $\check{{f}}_{12}^\s$ are close to the values
of $f_{12}^\s$ after the initial flow \footnote{This assumption is
reasonable as integrating out the $c,1$ boson changes
$f_{12}^\s$ only to second order in $\tilde{S}_{\rm int}$ and all
couplings in $\tilde{S}_{\rm int}$ are themselves small.} this implies
that the extended Hubbard model \fr{Hamiltonian} with a half-filled
bonding band describes a $C1S2$ phase as long as $V_\parallel/U$ and
$V_\perp/U$ are sufficiently small.

\subsubsection{$C1S0$ Phase}
\label{C1S0}
Using the interpretation of \fr{4kbEffH} as the low-energy limit of a
Kondo-Heisenberg model, there is a second parameter regime, namely the
one corresponding to \emph{antiferromagnetic} exchange interaction
between the spin-chain and the 1DEG. Here it is known that the RG flow
is towards a strong coupling phase in which both spin bosons become
gapped~\cite{zachar}. This phase occurs when $\check{{f}}_{12}^\s<0$.
following through the same arguments as in the $C1S2$ case, we
conclude that the resulting $C1S0$ phase occurs when $V_\parallel/U$,
$V_\perp/U$  are sufficiently large. In other words,  the Coulomb
interactions should not be screened too strongly in order for the
$C1S0$ phase to exist. 

Next we turn to the characterization of the physical properties of the
$C1S0$ phase. In this we are guided by the existing field theory
\cite{zachar,berg} and  numerical \cite{berg} studies of the KH
model. In particular it is known that the KH model exhibits
unconventional finite-wavevector pairing~\cite{berg}.  In terms of the
field theory the $C1S0$ phase is characterized by \cite{zachar} 
\bea
\langle\cos\big(\sqrt{8\pi}\Phi_{1,c}\big)\rangle\neq 0\ ,\quad
\langle\cos\big(\sqrt{2\pi}(\Phi_{1,s}+\Phi_{2,s})\big)\rangle\neq 0\
,\quad
\langle\cos\big(\sqrt{2\pi}(\Theta_{1,s}-\Theta_{2,s})\big)\rangle\neq 0.
\label{expectationvalues}
\eea
Concomitantly $\Theta_{1,c}$, ($\Theta_{1,s}+\Theta_{2,s})$ and
$\Phi_{1,s}-\Phi_{2,s}$ are fluctuating fields, i.e. one-point functions
of vertex operators of these fields vanish and (appropriate) two-point
functions decay exponentially. Using the fact that the expectation
values \fr{expectationvalues} are non-zero and that the only remaining
gapless degree of freedom is the antibonding charge sector we can
establish the dominant quasi long range order in the C1S0 phase. To
this end we consider the following order parameters:

\noindent
(1) bonding charge density wave (bCDW)
\be
\mathcal{O}_{bCDW}(n) = 
\frac{1}{2} \sum_{\s=\uparrow,\downarrow}  \big(a_{1,n,\s}^\dagger+a_{2,n,\s}^\dagger\big)\big(a_{1,n,\s}+a_{2,n,\s}\big).
\ee
Bosonizing this at vanishing interactions gives
\be
\mathcal{O}_{bCDW}(x) \sim  a_0\sqrt{\frac{2}{\pi}} \partial_x
\Phi_{1,c} 
-\frac{1}{\pi}\sin(2k_bx+\sqrt{2\pi}\Phi_{1,c})\cos(\sqrt{2\pi}\Phi_{1,s})+\ldots
\ee
(2) charge density wave (CDW)
\bea
\mathcal{O}_{CDW}(n) &=&
\sum_{\s=\uparrow,\downarrow}
a^\dagger_{1,n,\s}a_{1,n,\s}+a^\dagger_{2,n,\s}a_{2,n,\s}\nn
&\sim& a_0\sqrt{\frac{2}{\pi}}\partial_x(\Phi_{1,c}+\Phi_{2,c}) 
-\frac{1}{\pi}\cos\big(\sqrt{2\pi}\Phi_{1,s}\big)
\sin\big(2k_bx+\sqrt{2\pi}\Phi_{1,c}\big)\nn
&& -\frac{1}{\pi}\cos\big(\sqrt{2\pi}\Phi_{2,s}\big)
\sin\big(2k_{ab}x+\sqrt{2\pi}\Phi_{2,c}\big)+\nn
&&+ {\cal A} e^{i\sqrt{2\pi}(\Phi_{1,c}+\Phi_{2,c})}
\cos\big(\sqrt{2\pi}(\Theta_{1,s}-\Theta_{2,s})\big)
\cos\big(2(k_{ab}+k_{b})x\big)+
\ldots
\eea
where ${\cal A}$ is an amplitude which vanishes in the
$U\to0$ limit. This interaction induced terms for the charge
density wave operator are derived in Appendix~\fr{app:4kF}.
Using that that certain operators obtain expectation values in the 
$C1S0$ phase~\fr{expectationvalues}, we find the leading contribution
is
\be
{\cal O}_{CDW}(n)\bigg|_{C1S0} \sim \tilde{{\cal A}}\cos\big(2(k_b+k_{ab})n\big)e^{i\sqrt{2\pi}\Phi_{1,c}}  + \ldots
\ee
(3) d-wave superconductivity (SCd)
\bea
{\cal O}_{SCd}(n) &=& a_{1,n,\uparrow}a_{2,n,\downarrow}+a_{2,n,\uparrow}a_{1,n,\downarrow}\nn
&\sim& 2e^{i\sqrt{2\pi}\Theta_{1,c}}\cos(\sqrt{2\pi}\Phi_{1,s})
-2e^{i\sqrt{2\pi}\Theta_{2,c}}\cos(\sqrt{2\pi}\Phi_{2,s})\nn
&&+2e^{i\sqrt{2\pi}\Theta_{1,c}}\cos\big(2k_bx+\sqrt{2\pi}\Phi_{1,c}\big)
-2e^{i\sqrt{2\pi}\Theta_{2,c}}\cos\big(2k_{ab}x+\sqrt{2\pi}\Phi_{2,c}\big)
+\ldots
\eea
(4) antibonding pairing (abP) 
\bea
{\cal O}_{abP}(n) &=&(a^\dagger_{1,n,\uparrow}-a^\dagger_{2,n,\uparrow})
(a^\dagger_{1,n+1,\downarrow}-a^\dagger_{2,n+1,\downarrow})
-(a^\dagger_{1,n,\downarrow}-a^\dagger_{2,n,\downarrow})
(a^\dagger_{1,n+1,\uparrow}-a^\dagger_{2,n+1,\uparrow})\nn
&\sim&
{\cal A}_0
e^{-i\sqrt{2\pi}\Theta_{2,c}} \left[
  \cos\big(2k_{ab}(x+\frac{a_0}{2})+\sqrt{2\pi}\Phi_{2,c} \big)
+ \cos\big(\sqrt{2\pi}\Phi_{2,s}\big)\sin(k_{ab}a_0)\right]\nn
&+& e^{i\sqrt{2\pi}\Theta_{2,c}}\Bigg\{
\left[{\cal C}_1 \cos\left(\sqrt{4\pi}\Phi_{+,s}\right) - {\cal C}_3 \cos\left(\sqrt{4\pi}\Theta_{-,s}\right)\right] \cos\left(\sqrt{2\pi}\Phi_{1,c}+2k_bx\right)+\nn
&& \left[{\cal C}_2\cos\left(\sqrt{4\pi}\Phi_{+,s}\right) - {\cal C}_4 \cos\left(\sqrt{4\pi}\Theta_{-,s}\right) \right] \sin\left(\sqrt{2\pi}\Phi_{1,c}+2k_bx\right)\Bigg\}+\ldots
\label{OabP}
\eea
where the amplitudes ${\cal C}_a$ vanish in the $U\to 0$ limit,
$\Phi_{+,s} = (\Phi_{1,s}+\Phi_{2,s})/\sqrt{2}$ and 
$\Theta_{-,s}=(\Theta_{1,s}-\Theta_{2,s})/\sqrt{2}$. 
The interaction-induced contribution in the bosonized expression
\fr{OabP} is derived in Appendix \ref{app:abP}.
Using that some of the operators occurring in \fr{OabP} have non-zero expectation
values in the $C1S0$ phase \fr{expectationvalues}, we conclude that
the leading contribution is
\be
{\cal O}_{abP}(n)\bigg|_{C1S0} \sim (-1)^n  \tilde{\cal C}
e^{i\sqrt{2\pi}\Theta_{2,c}}+\ldots
\label{abPC1S0}
\ee
The bosonized form \fr{abPC1S0} of ${\cal O}_{abP}(n)$ coincides with
the PDW order parameter identified by Berg et. al. in the low-energy
description of the KHM~\cite{berg}, and with the analogous oder
parameter ${\cal O}_{PDW}$ proposed by Jaefari and Fradkin for the
doped two-leg ladder~\cite{FradkinArx}.

Using the bosonized expressions of the various order parameters
together with \fr{expectationvalues} we obtain the following results
for the long-distance asymptotics of correlation functions in the
$C1S0$ phase  
\bea
\langle {\cal O}_{CDW}(x)\ {\cal O}^\dagger_{CDW}(0)\rangle
&\propto& x^{-2}+\cos\big(2(k_{b}+k_{ab})x\big)\frac{{\cal A}}{|x|^{K_{2,c}}}+\ldots,\nn 
\langle {\cal O}_{bCDW}(x)\ {\cal  O}^\dagger_{bCDW}(0)\rangle
&\propto& e^{-|x|/\xi_b}~~({\rm  at~2}k_b)+\ldots,\nn 
\langle {\cal O}_{SCd}(x)\ {\cal  O}^\dagger_{SCd}(0)\rangle
&\propto&
\cos(2k_{ab}x)\frac{1}{|x|^{K_{2,c}}}\frac{1}{|x|^{1/K_{2,c}}}
+\ldots\ ,\nn
\langle {\cal O}_{abP}(x)\ {\cal
  O}^\dagger_{abP}(0)\rangle&\propto&
\frac{(-1)^{x/a_0}}{|x|^{1/K_{2,c}}}+\ldots
\ , \label{asyOabP}
\eea
where $\xi_b$ is correlation length for the bonding charge boson
and $K_{2,c}$ is the Luttinger parameter for the charge sector of the
antibonding band. These results suggest that there are two different
regimes:
\begin{enumerate}
\item{$K_{2,c}<1$}

Here the slowest decay of correlations is between the $2k_{ab}+2k_b$ components
of ${\cal O}_{CDW}$. Hence the C1S0 phase is identified as an
incommensurate charge density wave. 
\item{$K_{2,c}>1$}

Here the slowest decay of correlations is between the staggered components
of ${\cal O}_{abP}$ and concomitantly the C1S0 phase exhibits
unconventional fluctuation superconductivity with finite wavenumber
pairing. This ``pair-density wave'' phase was identified in
\cite{FradkinArx}. 

\end{enumerate}
Which regime is realized depends on the precise values of the
microscopic parameters $V_\perp$, $V_\parallel$. Integration of the RG
equations \fr{4kbRG} suggests that both regimes of $K_{2,c}$ can be
realized, although $K_{2,c}<1$ seems to be the more generic case.

As we mentioned before, the above analysis pertains to the case in which 
the Umklapp interaction is present automatically as a consequence of
the bonding band being half-filled. In the case when the Umklapp
interaction is induced through an external periodic potential, we
expect the same physics to emerge at low energies and in particular
both $C1S2$ and $C1S0$ phases to exist. 

\subsection{Chain Representation}
We now consider the effects of the $4k_b$ Umklapp interaction in the
chain representation. In order to simplify the analysis we will
focus on the case of extended density-density interactions along the
chains, which have the effect of decreasing the value of $K_c$ (see
the discussion at the beginning of section \ref{sec:chain}).
The low energy projection of the Umklapp term is
\be
H_W = \lambda\int dx \left[
ie^{i\sqrt{4\pi K_c}\Phi_c}
\left(\xi_R^1-i\xi^R_2\right)\left(\xi_L^1+i\xi_L^2\right)+{\rm h.c.} \right],
\ee
where we have rescaled the boson field $\Phi_c$ to absorb the
Luttinger parameter in the kinetic term of the Hamiltonian. The
perturbation $H_W$ has scaling dimension $d=1+K_c<2$ (for generic
repulsive interactions) and so this term is relevant in the renormalization
group sense. For long-range Coulomb interactions along the chains the
Luttinger parameter becomes small $K_c\ll1$ and this term is
\emph{strongly} relevant in the RG sense. It will therefore dominate 
the marginal four-fermion interactions in~\fr{FT} and should be
treated first. The Umklapp term is simplified by combining
the Majorana fermions into a complex (Dirac) fermion according to
$\mathcal{R}=(\xi_1^R+i\xi_2^R)/\sqrt{2}$ and
$\mathcal{L}^\dagger=(\xi_L^1-i\xi_L^2)/\sqrt{2}$ and  
then bosonizing $\mathcal{R},\mathcal{L}$ in terms of a Bose
field $\bar{\Phi}$ and its dual field $\bar{\Theta}$ following
Ref.\cite{GNT}. This gives 
\be
H_W = 
\frac{2\lambda}{\pi}\int dx\
\cos\left[\sqrt{4\pi}(\sqrt{K_c}\Phi_c+\bar{\Phi})\right].
\ee
We proceed by carrying out a canonical transformation
\bea
\Phi_\pm &=&
\frac{1}{\sqrt{2}}\left(\sqrt{K_c}\Phi_c\pm\bar{\Phi}\right),\qquad
\Theta_\pm=\frac{1}{\sqrt{2}}\left(\frac{\Theta_c}{\sqrt{K_c}}\pm
\bar{\Theta}\right),
\eea
where $\Theta_c$ is the field dual to $\Phi_c$. In terms of the new
bosonic fields the Hamiltonian density can be written as 
\bea
{\cal H} &=& \frac{v}{2}\left[K
  \left(\partial_x\Theta_+\right)^2+
K^{-1}\left(\partial_x\Phi_+\right)^2\right]+
m\cos\left(\sqrt{8\pi}\Phi_+\right)\nn  
&&+ \frac{v}{2}\left[K\left(\partial_x\Theta_-\right)^2
+K^{-1}\left(\partial_x\Phi_-\right)^2\right]
+g_1\partial_x\Theta_+\partial_x\Theta_-
+g_2\partial_x\Phi_+\partial_x\Phi_-  +
 \frac{iv_c}{2}\left(\xi_L^3\partial_x\xi_L^3-\xi_R^3\partial_x\xi_R^3\right)+\nn 
&& +\frac{iv_s}{2}\sum_a\left(\chi_L^a\partial_x\chi_L^a-\chi_R^a\partial_x\chi_R^a\right) - 2g_{\sigma-}\left(\xi_R^3\xi_L^3\right)\sum_a\left(\chi_R^a\chi_L^a\right)-2g_{\sigma+}\sum_{a>b}\left(\chi_R^a\chi_L^a\right)\left(\chi_R^b\chi_L^b\right)\nn
&& -\left(\xi_R^1\xi_L^1-\xi_R^2\xi_L^2\right)\left[2g_{c,ss}\left(\xi_R^3\xi_L^3\right)+g_{\rho,-}\left(\xi_R^1\xi_L^1-\xi_R^2\xi_L^2\right)+2g_{c,st}\sum_a\left(\chi_R^a\chi_L^a\right)\right],
\eea
where $g_{1,2}$ and $m$ are redefined coupling constants and 
\be
v=\frac{1}{2}\sqrt{\big(v_c+\tilde{u}/K_c\big)
\big(v_c+\tilde{u}K_c\big)}\ ,\quad
K^2=K_c\frac{\tilde{u}+v_cK_c}{v_c+\tilde{u}K_c}.
\ee
As we are considering strongly repulsive interactions we have $K\ll
1$. By construction the cosine term in the sine-Gordon model for the
$\Phi_+$ boson is strongly relevant and will reach strong coupling
before any of the other running couplings becomes large. In other
words, the Umklapp-induced gap in the $\Phi_+$ sector will be large
compared to all other low-energy scales.

In the next step we want to integrate out the ${\Phi}_+$ boson, 
similarly to what we did in the band representation. To this end we
express the $\xi^{1,2}$ Majorana fermions in terms of the Dirac
fermions ${\cal R}$ and ${\cal L}$ and then proceed to bosonize
them. The four-fermion interactions that involve the $\xi^{1,2}$
Majorana fermions are proportional to
\bea
\left(\xi_R^1\xi_L^1-\xi_R^2\xi_L^2 \right) &=& \frac{1}{2}\left[\left(\xi_R^1+i\xi_R^2\right)\left(\xi_L^1+i\xi_L^2\right)+\left(\xi_R^1-i\xi_R^2\right)\left(\xi_L^1-i\xi_L^2\right)\right],\nn
&=& \mathcal{R}^\dagger\mathcal{L}^\dagger+\mathcal{R}\mathcal{L}
\sim \frac{i}{2\pi}\cos\left[\sqrt{2\pi}\left(\Theta_++\Theta_-\right)\right].
\eea
When integrating out the $\Phi_+$ boson we therefore only
generate interactions proportional to $\cos\big(\sqrt{8\pi}{\Theta}_-\big)$,
which are irrelevant as $K\ll 1$. At energies small compared to the
mass gap of the $\Phi_+$ boson, the effective Hamiltonian density has
the form  
\bea
{\cal H}_\mathrm{eff} &=& \frac{\tilde{v}}{2} \left[ \tilde{K}
  \left(\partial_x\Theta_-\right)^2+\tilde{K}^{-1}
  \left(\partial_x\Phi_-\right)^2\right] +
\frac{iv_c}{2}\left[\xi_L^3\partial_x\xi_L^3-\xi_R^3\partial_x\xi_R^3\right]
+\nn 
&& \frac{iv_s}{2}\left[\chi_L^a\partial_x\chi_L^a-\chi_R^a\partial_x\chi_R^a\right] - 2\tilde{g}_{\sigma-}\left(\xi_R^3\xi_L^3\right)\sum_a\left(\chi_R^a\chi_L^a\right) - 2\tilde{g}_{\sigma+}\sum_{a>b}\left(\chi_R^a\chi_L^a\right)\left(\chi_R^b\chi_L^b\right).
\label{eq:Heff4kF+2Q}
\eea
where $\tilde{g}$ are renormalized couplings, $\tilde{v}$ is the
renormalized velocity and $\tilde{K}$ is the renormalized Luttinger
parameter. The effective Hamiltonian~\fr{eq:Heff4kF+2Q} is remarkably
similar in form to the field theory limit of the Kondo-Heisenberg
model, with the difference that the velocity of the singlet and
triplet Majorana modes are not equal.  

In order to analyze the effective theory \fr{eq:Heff4kF+2Q} further we
carry out a renormalization group analysis, which gives
\be
\dot{\tilde{g}}_{\s-} = -\frac{2}{\pi v_s}\tilde{g}_{\s-}\tilde{g}_{\sigma+}\ ,
\qquad \dot{\tilde{g}}_{\sigma+} = -\frac{\tilde{g}_{\s-}^2}{\pi v_c}
-\frac{\tilde{g}_{\sigma+}^2}{\pi v_s}\ . 
\label{RGchain1}
\ee
These RG equations are easily integrated. Defining 
$g_\pm=\frac{\tilde{g}_{\sigma-}}{\pi\sqrt{v_cv_s}}\pm
\frac{\tilde{g}_{\sigma+}}{\pi v_s}$, Eqs.~\fr{RGchain1} become
$\dot{g}_\pm=\mp g^2_\pm$, which have the solution
\be
g_\pm(l)=\frac{g_\pm(l_0)}{1\pm g_\pm(l_0) (l-l_0)}.
\ee
Assuming that $g_{\sigma,\pm}$ renormalize only weakly from their bare
values up to the RG time $l_0$ at which the $\Phi_+$ sector reaches
strong coupling, we conclude that
\be
\tilde{g}_{\sigma,\pm}(l_0)>0.
\ee
This then implies that the RG flow of $g_+$ is always towards weak
coupling. On the other hand, $g_-$ flows to a strong coupling C1S0
fixed point if
\be
\tilde g_{\sigma-}(l_0)>\tilde g_{\sigma+}(l_0)\sqrt{\frac{v_c}{v_s}},
\label{ineq}
\ee
In order to get a sense of what this requirement implies in terms of
the underlying microscopic theory we consider the case when $\tilde
g_{\sigma\pm}(l_0)$ are close to their bare values and
$U,V_\parallel,V_\perp$ are small. Then
\be
g_{\sigma-}\sim\frac{V_\perp a_0}{2}\ ,\quad
g_{\sigma+}\sim a_0\big(U+2\cos(2k_Fa_0)V_\parallel\big),
\ee
where $a_0$ is the lattice spacing and $k_F\approx\pi/2$.
\be
V_\perp\agt 2\sqrt{\frac{v_c}{v_s}}\big(U-2V_\parallel).
\ee
Hence, just as was the case for the weak-coupling analysis of the
previous subsection, having repulsive interactions between
neighboring sites is crucial for driving the systems into a C1S0
phase. Having established the existence of a C1S0 phase in the chain
representation, the next step would be to determine which correlations
are dominant. This is difficult for the following reason. General
local observables can be expressed in terms of Ising models, but it
remains an open problem to determine how products of Ising order and
disorder operators transform under Tsvelik's transformation
\fr{MajoanatrafoL}. 

\section{$3k_b+k_{ab}$ Umklapp}
\label{sec:3kb+kab}
In this section we consider the $3k_b+k_{ab}$ Umklapp process. 
Unlike in the $4k_b$ case, where the Umklapp emerged automatically for
a particular value of the doping as a result of the Hubbard
interaction, we now need to introduce an external periodic potential
with the appropriate modulation. 

\subsection{Chain Representation}
The $3k_b+k_{ab}$ Umklapp is most easily treated in the chain
representation. We add to the low-energy Hamiltonian~\fr{FT} the term
\bea
H_W &=& \lambda\int dx \left[\rho^{(-)}_{4k_{F,0}+Q}(x) + {\rm
    h.c.}\right]\nn
&=& -i\lambda\int dx
\big[(\cos\big(\sqrt{4\pi}\Phi_c\big)~\xi_R^1-\sin\big(\sqrt{4\pi}\Phi_c\big)~\xi_R^2)\xi_L^3
+
\xi_R^3(\cos\big(\sqrt{4\pi}\Phi_c\big)~\xi_L^1+\sin\big(\sqrt{4\pi}\Phi_c\big)~\xi_L^2)\big], 
\label{majumklapp}
\eea
The scaling dimension of $H_W$ is $d=1+K_c<2$ and the Umklapp is
therefore strongly relevant in the RG sense for the case of strong,
long-ranged repulsive interactions ($K_c\ll 1$), see
the discussion at the beginning of section \ref{sec:chain}. In this
case, the Umklapp term quickly flows to strong coupling under RG, while other
interactions remain small in comparison. However, a na\"ive mean-field
treatment of the Umklapp term is not possible as it would break a (hidden)
continuous $U(1)$ symmetry of the Hamiltonian. In order to analyze the
effects of $H_W$ we therefore perform a field redefinition (in the path
integral) 
\bea
&& \xi_R^1 = \cos\big(\sqrt{4\pi}\Phi_c\big) r + \sin\big(\sqrt{4\pi}\Phi_c\big) r_0,\qquad \xi_R^2 = -\sin\big(\sqrt{4\pi}\Phi_c\big) r + \cos\big(\sqrt{4\pi}\Phi_c\big) r_0,\nn
&& \xi_L^1 = \cos\big(\sqrt{4\pi}\Phi_c\big) l - \sin\big(\sqrt{4\pi}\Phi_c\big) l_0,\qquad \xi_L^2 = \sin\big(\sqrt{4\pi}\Phi_c\big) l + \cos\big(\sqrt{4\pi}\Phi_c\big) l_0.
\label{fieldtransform}
\eea
The new fields $r_0$, $l_0$, $r$, $l$ are fermionic in nature and the
Jacobian of \fr{fieldtransform} is unity. The
transformation~\fr{fieldtransform} diagonalizes the Umklapp 
interaction and removes from it the total charge boson
$\Phi_c$
\be
H_W = i\lambda(\xi_L^3r+l\xi_R^3).
\ee
The Lagrangian density then reads
\bea
\mathcal{L} &=& \frac{1}{8\pi}\big[v_c^{-1}(\partial_\tau
  \Phi)^2+v_c(\partial_x\Phi)^2\big] +
\sqrt{K_c}rr_0(\partial_\tau-iu\partial_x)\Phi
-\sqrt{K_c}ll_0(\partial_\tau+iu\partial_x)\Phi\nn 
&+&\frac{1}{2} r(\partial_\tau-iu\partial_x)r
+\frac{1}{2}r_0(\partial_\tau-iu\partial_x)r_0
+\frac{1}{2}l(\partial_\tau+iu\partial_x)l
+\frac{1}{2}l_0(\partial_\tau+iu\partial_x)l_0\nn 
&+& \frac{1}{2}\xi_R^3(\partial_\tau-i v_c \partial_x)\xi_R^3 +
\frac{1}{2}\xi_L^3(\partial_\tau+i v_c \partial_x)\xi_L^3 +
\frac{1}{2}\sum_{a=1}^3\big[\chi_R^a(\partial_\tau-iv_s\partial_x)\chi_R^a+\chi_L^a(\partial_\tau+iv_s\partial_x)\chi_L^a\big]\nn 
&+& i\lambda(\xi_L^3r+l\xi_R^3) + V_{\rm int},
\label{Lagrangian}
\eea
where we have defined $\Phi=\sqrt{4\pi/ K_c}\Phi_c$ and 
\bea
V_{\mathrm{int}} &=& -2g_{s,cc}(\xi_R^3\xi_L^3)(rl-r_0l_0)-g_{\rho,-}(rl-r_0l_0)^2\nn
&-& 2\big[g_{c,st}(rl-r_0l_0)+g_{\sigma,-} (\xi_R^3\xi_L^3)\big]
\sum_a(\chi_R^a\chi_L^a)-2g_{\sigma,+}\sum_{a>b}(\chi_R^a\chi_L^a)(\chi_R^b\chi_L^b). 
\eea
To make further progress we now drop the terms containing
$rr_0\partial\Phi$ and $ll_0\partial\Phi$. These terms carry non-zero
Lorentz spin and do not produce singularities in perturbation
theory. We also note that the corresponding interaction vertices do
not induce a mass for the $r_0$ or $l_0$ fermions.

Inspection of \fr{Lagrangian} then indicates that the Umklapp interaction
acts as a mass term for the fermions $(r,\xi_L^3)$ and $(l,\xi_R^3)$
and the neglected terms renormalize these gaps, in accordance with the
scaling dimension of the original $H_W$. These substantial gaps
allow us to integrate out the Fermi fields
Fermi fields $r,l,\xi_{R,L}^3$, leading to the following effective
theory at low energies
\bea
{\cal H}_{\rm{eff}} &=& {\cal H}_c + i\frac{u}{2}(l_0\partial_xl_0-r_0\partial_xr_0)+\frac{iv_s}{2}\sum_{a}(\chi_L^a\partial_x\chi_L^a-\chi_R^a\partial_x\chi_R^a) \nn
&+&2\tilde{g}_{c,st} (r_0l_0)\sum_a(\chi_R^a\chi_L^a) -
2\tilde{g}_{\sigma+}\sum_{a>b}(\chi_R^a\chi_L^a)(\chi_R^b\chi_L^b). 
\label{effectiveH}
\eea
This effective Hamiltonian is of the same form as \fr{eq:Heff4kF+2Q},
found in the analysis of the $4k_F+2Q$ Umklapp, so it also is similar
to the Kondo-Heisenberg model. If the four-fermion couplings are
large, such that we can perform a mean-field treatment, the resulting
theory is a C1S0 phase, where the charge boson $\Phi_c$ remains
massless, whilst the $r_0,\ l_0$ and $\chi$ Majorana fermions have
dynamically generated masses. To extract the low-energy behavior of
our effective Hamiltonian with weak four-fermion coupling, let us
consider the RG equations 
\bea
\dot{\tilde{g}}_{c,st} &=& -\frac{2}{\pi v_s}\tilde{g}_{c,st}\tilde{g}_{\sigma+},\\
\dot{\tilde{g}}_{\sigma+} &=& -\frac{\tilde{g}_{c,st}^2}{\pi u} -\frac{\tilde{g}_{\sigma+}^2}{\pi v_s}.
\eea
These equations can be integrated in the same way as \fr{RGchain1}.
The RG flow is towards a C1S0 strong coupling phase if
\be
\tilde{g}_{c,st}(l_1) > \tilde{g}_{\sigma+}(l_1)\sqrt{\frac{u}{v_s}},
\label{criterion}
\ee
where $l_1$ is the RG time at which the Umklapp interaction strength
$\lambda$ reaches strong coupling. Considering the case when the
renormalized couplings are close to their original values we find that
\fr{criterion} is generically satisfied as for repulsive interactions
$v_s<v_c$. 

In summary, depending on the values of the coupling constants the
effective Hamiltonian~\fr{effectiveH} describes either a C1S2 or a
C1S0 phase. When the criterion~\fr{criterion} is not met, the
effective Hamiltonian flows to weak-coupling under RG and we end up in
a C1S2 phase, where only the antisymmetric charge boson obtains a
mass. Pairing fluctuations may occur with finite-wavevector, but the
correlations are unlikely to be dominant in the absence of a spin gap. 
On the other hand, if \fr{criterion} is fulfilled there is a spin gap
and it is tempting to speculate that at low energies strong
superconducting correlations exist. The determination of the
long-distance asymptotics of local operators in this C1S0 phase is
difficult, because their field theory expressions generally involve
Ising order and disorder operators and it is not known how these
transform under \fr{MajoanatrafoL}.  
\subsection{Band Representation}
In the band representation the $3k_b+k_{ab}$ Umklapp scattering adds a
term to the Hamiltonian~\fr{eqs:FieldHamiltonian} of the form
\be
H_W =\tilde \lambda\int dx~(I^L_{11})^\dagger I^R_{12}+(I^L_{12})^\dagger
I^R_{11}+{\rm H.c.}
\ee
In the absence of the Umklapp interaction, the one-loop renormalization
group equations have been derived in \cite{balents,CCL}. The additional
terms in the one-loop RG equations are most easily derived using
operator product expansions. The one-loop RG equations are found to be
of the form
\bea
\dot{c}^\rho_{11}&=&
-\frac{\alpha}{4}\left[
(c^\rho_{12})^2+3(c^\sigma_{12})^2\right]+2\lambda^2\ ,\nn
\dot{c}^\rho_{12}&=&-\frac{1}{4}\left[
c^\rho_{11}c^\rho_{12}+3c^\sigma_{11}c^\sigma_{12}\right]
-\frac{1}{4}\left[
c^\rho_{12}c^\rho_{22}+3c^\sigma_{12}c^\sigma_{22}\right]
+\frac{1}{2}\left[c^\rho_{12}f^\rho_{12}+3c^\sigma_{12}f^\sigma_{12}\right]
+\lambda^2\ ,\nn
\dot{c}^\rho_{22}&=&-\frac{\alpha}{4}\left[
(c^\rho_{21})^2+3(c^\sigma_{21})^2\right]\ ,\nn
\dot{c}^\sigma_{11}&=&-(c^\sigma_{11})^2
-\frac{\alpha}{2}c^\sigma_{12}(c^\rho_{12}+c^\sigma_{12})\ ,\nn
\dot{c}^\sigma_{12}&=&-\frac{1}{4}\left[(c^\rho_{11}+
c^\rho_{22})c^\sigma_{12}+(c^\rho_{12}+2c^\sigma_{12})
(c^\sigma_{11}+c^\sigma_{22})\right]
+\frac{1}{2}\left[c^\rho_{12}f^\sigma_{12}+c^\sigma_{12}f^\rho_{12}
-2c^\sigma_{12}f^\sigma_{12}\right]\ ,\nn
\dot{c}^\sigma_{22}&=&-(c^\sigma_{22})^2
-\frac{\alpha}{2}c^\sigma_{12}(c^\rho_{12}+c^\sigma_{12})\ ,\nn
\dot{f}^\rho_{12}&=&\frac{1}{4}\left[
(c^\rho_{12})^2+3(c^\sigma_{12})^2\right]+\lambda^2\ ,\nn
\dot{f}^\sigma_{12}&=&-(f_{12}^\sigma)^2+\frac{1}{2}
c^\sigma_{12}(c_{12}^\rho-c_{12}^\sigma)\ ,\nn
\dot{\lambda}&=&\frac{\lambda}{2}\left[c_{11}^\rho+c_{12}^\rho+f_{12}^\rho
\right]. \nonumber
\label{RG3kbkab}
\eea
where $\alpha=(v_1+v_2)/4v_1v_2$ and the coupling constants have been rescaled
according to
\be
c_{ij}=\frac{\tilde{c}_{ij}}{\pi(v_i+v_j)}\ ,\quad
f_{ij}=\frac{\tilde{f}_{ij}}{\pi(v_i+v_j)}\ ,\quad
\lambda=\frac{\tilde{\lambda}}{\sqrt{2\pi v_1 \pi(v_1+v_2)}}.
\ee

The next step is then to numerically integrate \fr{RG3kbkab}
in an attempt to infer the strong-coupling fixed point. 
To be explicit, let us consider a particular example at vanishingly
weak coupling, when the the $3k_b+k_{ab}$ Umklapp interaction emerges
at a particular band filling. In the absence of interactions the Fermi  
momenta of bonding/antibonding bands are
\be
k_{b}={\rm arccos}\left(-\frac{t_\perp+\mu}{2t}\right)\,\quad
k_{ab}={\rm arccos}\left(\frac{t_\perp-\mu}{2t}\right)\ .
\ee
For the Umklapp to be present as a result of the Hubbard interactions
we require $3k_b+k_{ab}=2\pi$. For the ladder with $2t_\perp=t$ this
corresponds to a chemical potential of $\mu = -0.245898t$, resulting
in $ v_b = 1.98380ta_0$, $v_{ab}=1.85570ta_0$, and concomitantly
$\alpha= 1.0011$. Integrating the RG equations leads to a flow with
$f_{12}^\sigma\rightarrow0$, $c_{12}^\rho\to\infty$ and
\bea
&&c_{11}^\rho\rightarrow-\frac{1}{2}c_{12}^\rho\ ,\quad
c_{22}^\rho\rightarrow-\frac{1}{2}c_{12}^\rho\ ,\quad
f_{12}^\rho\rightarrow\frac{1}{2}c_{12}^\rho\ ,\nn
&&c_{11}^\sigma\rightarrow -c_{12}^\rho\ ,\quad
c_{22}^\sigma\rightarrow -c_{12}^\rho\ ,\quad
c_{12}^\sigma\rightarrow c_{12}^\rho\ .
\eea

In the case when $U=8V_\parallel=16V_\perp$ and Umklapp coupling
$\tilde\lambda=U$, the renormalization group flow is $f_{12}^\rho\rightarrow\infty$
while
\be
\frac{c_{ij}^\sigma}{f_{12}^\rho}\rightarrow 0\ ,\quad
\frac{f_{12}^\sigma}{f_{12}^\rho}\rightarrow 0\ ,\quad
\frac{c_{11}^\rho}{f_{12}^\rho}\rightarrow0.9869\ ,\quad
\frac{c_{12}^\rho}{f_{12}^\rho}\rightarrow 0.1648\ ,\quad
\frac{c_{22}^\rho}{f_{12}^\rho}\rightarrow -0.006568\ ,\quad
\frac{\lambda}{f_{12}^\rho}\rightarrow 0.7169.
\ee
Provided the extended interactions are sufficiently weak,
we find the same pattern of diverging couplings, but the
final ratios depend on $v_{1,2}$.
In the band representation it is difficult to analyze the fixed point
Hamiltonian further and we leave this for future studies.

\section{Numerical Results: DMRG}
\label{sec:DMRG}
In this section we use the density matrix renormalization
group (DMRG) algorithm \cite{dmrg1,dmrg2} to study the extended
Hubbard model on the two-leg ladder. Hubbard-like models have been
previously studied using DMRG, both on single chains and multiple leg
ladders~\cite{noack,NoackBulutScalapinoZacher97,eric,WhiteAffleckScalapino02,noack94,endres,weihong,orignac}. 
In the following we first consider the case where the Umklapp interaction 
does not play a role and analyze the resulting ``generic strong
coupling regime'' in Section~\ref{gscr}. Having established this crucial
reference point, we then turn to the case where the Umklapp interaction
is marginally relevant.

\subsection{Generic Strong Coupling Regime}
\label{gscr}
For sufficiently small extended interactions, the (weak-coupling)
renormalization group flow of the model is towards a strong-coupling
fix point described by a SO(6) Gross-Neveu
model~\cite{Fabr93,schulz,balents,lin}, which can be analyzed by exact
methods~\cite{ek07}. In this theory three of the bosons, $\Theta_{-,c}$,
$\Phi_{+,s}$ and $\Phi_{-,s}$, become massive under the RG flow whilst the
remaining massless charge boson $\Phi_{+,c}$ is described by a
U(1) Luttinger liquid theory. The values to which the bosons become pinned 
by the RG flow can be extracted from a classical analysis of the
effective theory. Following such an analysis, the asymptotic form of
the two-point function of the order parameters discussed in Section~\ref{C1S0}
 are found to be~\cite{schulz,balents}
\bea
\langle {\cal O}_{CDW}(x)\ {\cal O}^\dagger_{CDW}(0)\rangle_{\rm
  Generic}&\propto& {\cal A}_1x^{-2}+{\cal A}_2\cos\big(2(k_b+k_{ab})x\big)x^{-2K_c},\nn 
\langle {\cal O}_{SCd}(x) {\cal O}^\dagger_{SCd}(0)\rangle_{\rm Generic}&\propto& |x|^{-1/2K_{c}},\nn
\langle {\cal O}_{abP}(x) {\cal O}^\dagger_{abP}(0)\rangle_{\rm Generic}&\propto& |x|^{-1/2K_{c}},
\label{genericops}
\eea
where $K_{c}$ is the Luttinger parameter for the remaining massless
$\Phi_{+,c}$ boson. The $2k_F$ response of the CDW and bCDW order parameters
are blocked by the presence of a spin gap, as is discussed in Appendix~\ref{app:2kF}. 
The second term in the two-point function of the charge density wave (CDW) order
parameter is interaction-induced, with the amplitude ${\cal A}_2$ vanishing in the 
$U\to0$ limit; further discussion of interaction-induced terms may be found in 
Appendix~\ref{app:4kF}.

As an example of the generic strong coupling regime, we present results for
the Hamiltonian~\fr{Hamiltonian} on the $64\times2$ ladder with $t=t_\perp=1$, $U=4$ 
and $V_\parallel=V_\perp=W_{1,2}=0$. As is usual with DMRG calculations, we
take open boundary conditions on the ends of the ladder~\cite{dmrg2}. 
We consider the system with $N=110$ electrons and keep up to $m=1500$ 
density matrix states in the DMRG simulation, leading to 
truncation errors of $\sim3\times10^{-6}$. Performing an extrapolation
of the ground state energy per site against the number of density matrix states kept in the
calculation allows one to estimate the relative error in quantities calculated by the DMRG algorithm.
We define the relative error in the ground state energy per site $\epsilon = (\bar{E}_0-\bar{E}_{\rm DMRG})/\bar{E}_{0}$,
where $\bar{E}_0$ is the extrapolated value and $\bar{E}_{\rm DMRG}$ is the
measured value for the ground state energy per site. In this case, we find that 
$m=1500$ density matrix states results in a relative error of $\epsilon\approx5\times10^{-4}$.

\begin{figure}[ht]
\begin{tabular}{ccc}
(a)\includegraphics[width=0.46\textwidth]{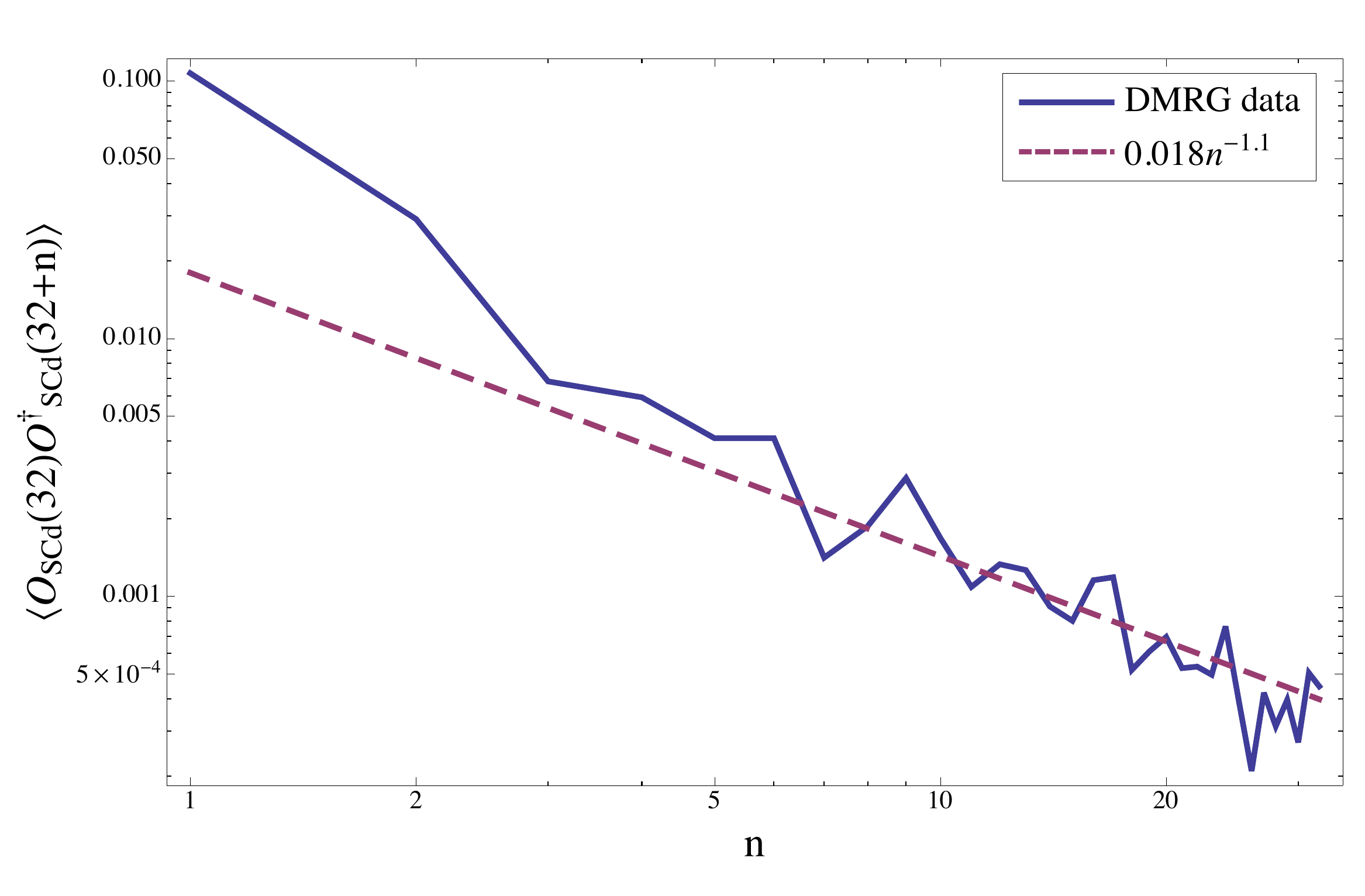} &\quad&
(b)\includegraphics[width=0.46\textwidth]{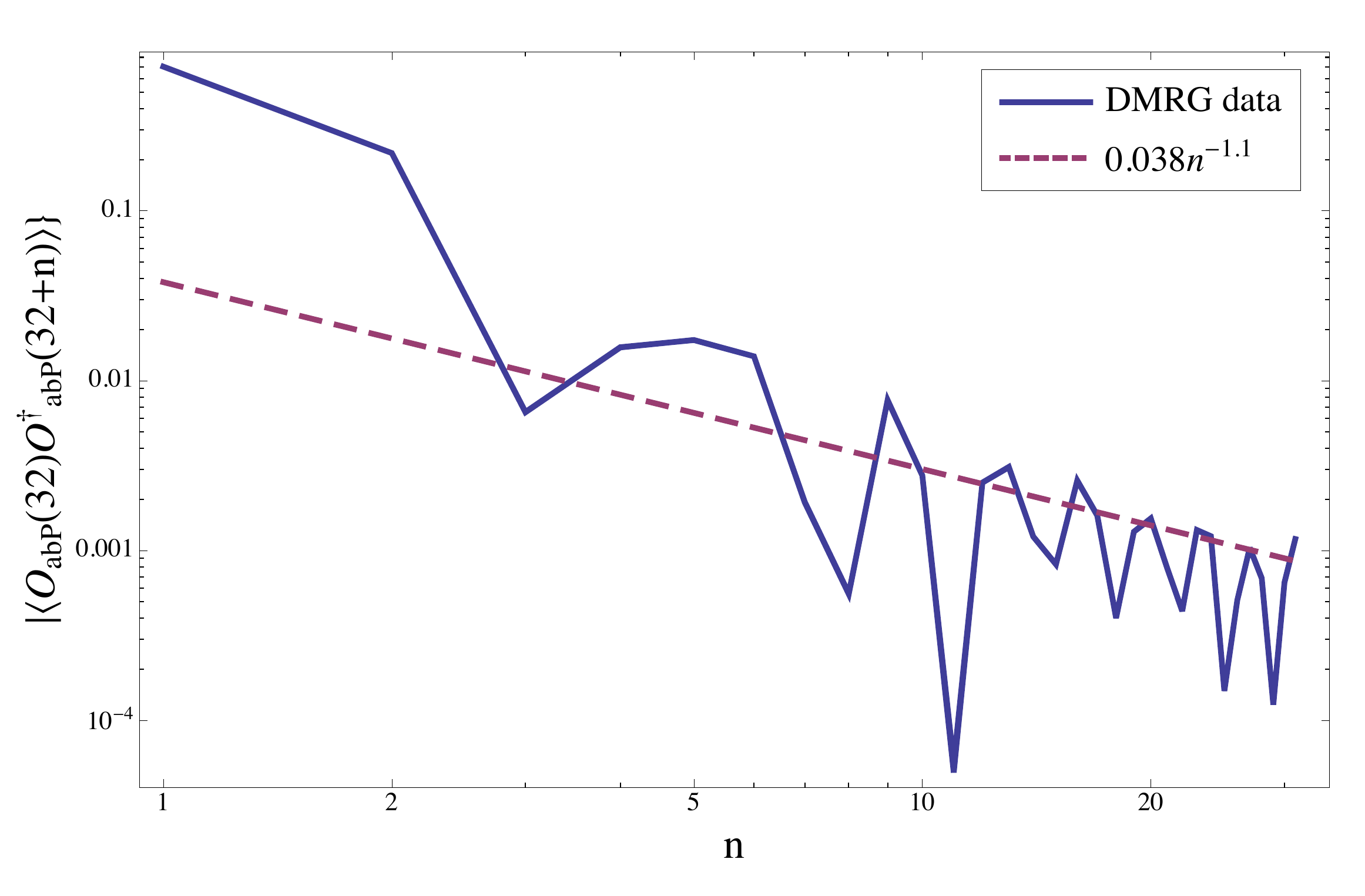}
\end{tabular}
\caption{DMRG data (solid) and power law fits (dashed) for 
(a) the superconducting d-wave order parameter ${\cal O}_{SCd}$, 
and (b) the antibonding pairing order parameter ${\cal O}_{abP}$ 
on the $64\times2$ ladder with 
$t=t_\perp=1$, $U=4$, $V_\parallel=V_\perp=0$ and $N=110$ electrons.
Oscillations present in both two-point functions are contributions
from sub-leading terms. 
 }
\label{fig:generic}
\end{figure}

Figure~\ref{fig:generic} shows the calculated two-point functions of the SCd and abP order
parameters and appropriate power law fits. Additional oscillations at $2k_{ab}$ are
observed in the two-point function of the antibonding pairing order parameter, 
which may be due to a small amplitude for the power law decay term and/or
a large spin-correlation length for the exponentially decaying terms. This would
be consistent with a small spin gap in the system. The power law fits to the two-point
functions give the Luttinger parameter for the massless $\Phi_{+,c}$ boson as
 $K_{c}\approx0.45$. 
 
\begin{figure}
\begin{center}
\includegraphics[width=0.65\textwidth]{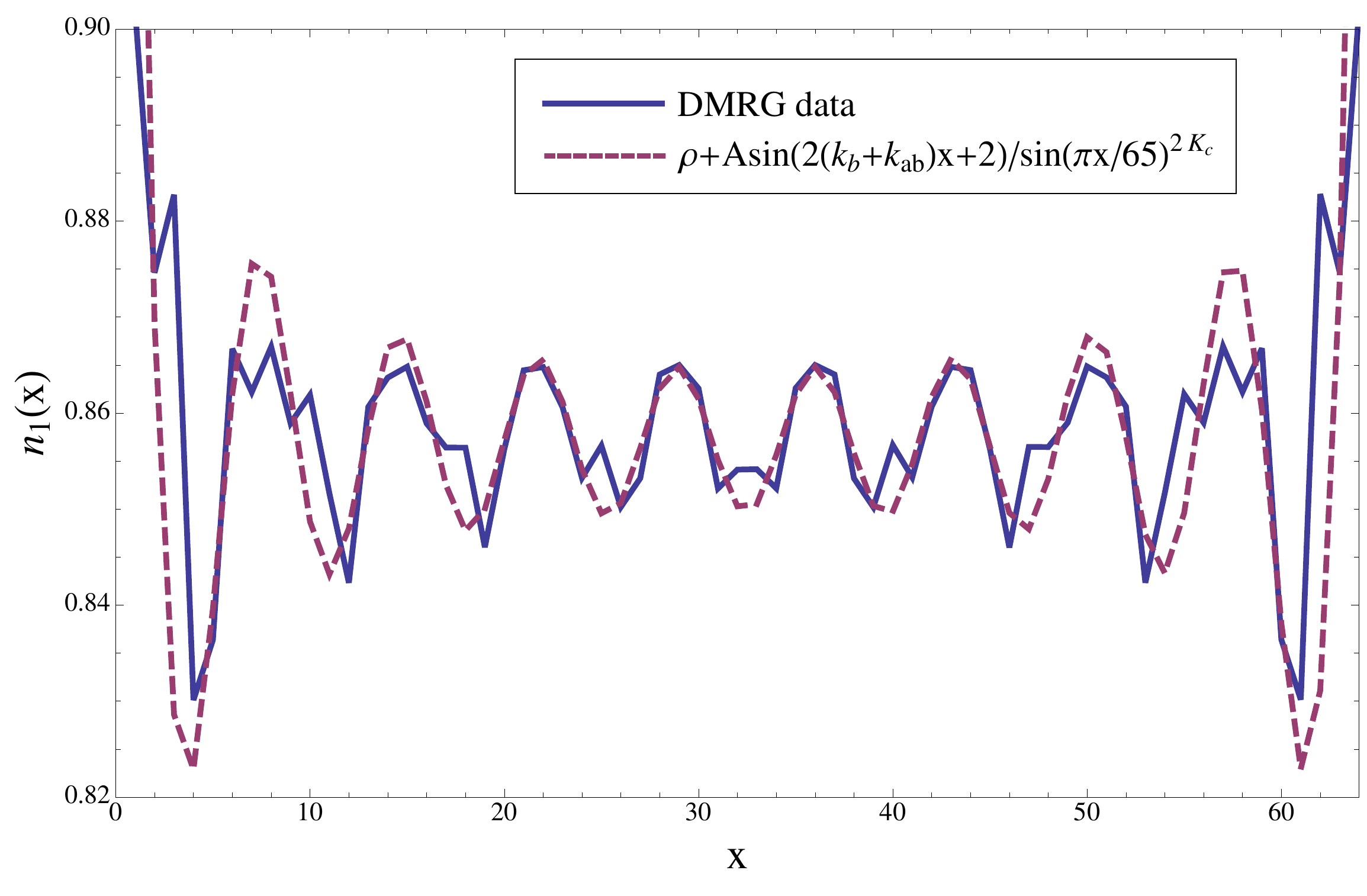}
\end{center}
\caption{DMRG data (solid) and fit (dashed) for the one-point
function of the density operator on leg $1$ of the  
the $64\times2$ ladder with $t=t_\perp=1$, $U=4$, $V_\parallel=V_\perp=0$ 
and $N=110$ electrons. The fit function parameters take values $\rho = 0.857$, 
$A=0.0054$ and $2K_c = 0.91$. The bonding and antibonding wavevectors are
given by $k_b=\rho_{b}\pi/2=1.17\pi/2$ and $k_{ab}=\rho_{ab}\pi/2=0.55\pi/2$,
with $\rho_{b}$ ($\rho_{ab}$) the average density in the bonding (antibonding) band.
The fit function takes the form discussed in
Ref.~\cite{Eggert09} for the $4k_F$ Wigner crystal oscillations. Additional oscillations arise from
the sub-leading contributions of the charge density}
\label{fig:genericdens}
\end{figure}

 Figure~\ref{fig:genericdens} show the one-point function of the density operator 
 across leg-$1$ of the ladder. The oscillations in the density are induced by the 
 the open boundary conditions on the ends of the ladder. The presence of a spin
 gap in the system suppresses the $2k_F$ response (Friedel oscillations) in the ladder,
 consequently the leading order oscillations occur at $4k_F=2(k_{b}+k_{ab})$, known
 as ``Wigner crystal'' oscillations~\cite{Eggert09}. We fit the ``Wigner crystal'' oscillations
 to the standard form~\cite{Eggert09}
 \be
 \langle n(x) \rangle_{4k_F} = \rho + A \frac{\sin\left(4k_Fx+\varphi\right)}{\sin\left(\frac{\pi}{L+1}x\right)^{2K_c}},
 \ee
 where $A$ and $\varphi$ are fitting parameters, $\rho$ is the average electron density
 and $L$ is the length of the ladder. Additional oscillations which arise in the one-point function
 of the density operator are from the sub-leading contributions to the density operator, such
 as those discussed in Appendix~\ref{app:4kF}. In the presented fit we use the value for the Luttinger
 parameter extracted from the two-point functions of the SCd and abP order parameter.
 The value of the Luttinger parameter is also consistent with the long-distance asymptotics
 of the two-point function of the charge density operator, as would be expected from the
 analysis of the one-point function.

It is clear that the dominant correlations for the discussed generic strong coupling regime
depend upon the microscopic parameters of the Hamiltonian~\fr{Hamiltonian}. For the case
which we have considered, the Luttinger parameter $K_c<1/2$ and the phase is best described
by charge density wave correlations, with the leading contribution arising from the $2(k_b+k_{ab})$
interaction-induced component of the charge density.
\subsection{$4k_b$ Umklapp}
As is discussed in detail in Section~\ref{sec:4kb}, there
are two possible phases when the $4k_b$ Umklapp interaction
is present and marginally relevant for the considered initial conditions. 
We consider in turn the $C1S2$ phase and the $C1S0$ phase which may 
occur as a result of the $4k_b$ Umklapp modifying the renormalization
group equation. To that end we have carried out DMRG computations on
the Hamiltonian
\be
H=H_{\rm ladd}(\pi)+\mu_-\sum_{j,\s}\big(c^\dagger_{1,j,\s}-c^\dagger_{2,j,\s}\big)
\big(c_{1,j,\s}-c_{2,j,\s}\big),
\label{Hpi}
\ee
where $H_{\rm ladd}(K)$ is given by \fr{Hamiltonian} and the bonding
band is at quarter-filling. The additional term in \fr{Hpi}
corresponds to a chemical potential for the antibonding pair and is
introduced for convenience so that the antibonding density can be
varied while keeping the interaction parameters constant. A
quarter-filled bonding band requires an applied external potential of
wavevector $4k_b=\pi$ to activate the $4k_b$ Umklapp interaction. 

The reason for studying the model \fr{Hpi} rather than the doped
ladder with quarter filled bonding band but without external potential
is that in the latter both the Mott gap and spin gaps depend on the
interaction strengths $U$, $V_\parallel$, $V_\perp$ and therefore
cannot be tuned independently. As a result spin and charge gaps can be
comparable in size and small, which makes a numerical analysis
extremely challenging. In fact, our DMRG results for this case are
inconclusive in the sense that we have not found convincing evidence
for the existence of a C1S0 phase. 

On the other hand, applying an external potential as in \fr{Hpi}
allows us to control the Mott gap in the bonding sector without
significantly affecting spin gaps. A sizeable Mott gap makes the
numerical analysis much simpler.

\subsubsection{$C1S2$ Phase}
The RG analysis of section \ref{sec:4kb} shows that for sufficiently weak
extended interactions (small $V_\parallel,V_\perp$) the
renormalization group flow of the extended Hubbard model in the
presence of a $4k_b$ Umklapp interaction is towards a $C1S2$ fixed point.
The two-point functions of the order parameters discussed in
section~\ref{sec:4kb} then have the following forms
\bea
\langle {\cal O}_{CDW}(x)\ {\cal
  O}^\dagger_{CDW}(0)\rangle_{C1S2}&\propto& 
{\cal A}_3\cos(2k_{ab}x){|x|^{-K_{2,c}-K_{2,s}}}+{\cal A}_4 \cos(2k_bx)|x|^{-K_{1,s}} + {\cal A}_5x^{-2} +\ldots\ ,\nn
\langle {\cal O}_{bCDW}(x)\ {\cal O}^\dagger_{bCDW}(0)\rangle_{C1S2}&\propto& e^{-|x|/\xi_{b}} \quad({\rm at~2}k_b)+\ldots\ ,\nn
\langle {\cal O}_{SCd}(x) {\cal
  O}^\dagger_{SCd}(0)\rangle_{C1S2}&\propto& 
\cos(2k_{ab}x) |x|^{-K_{2,c}-1/K_{2,c}}+\ldots\ ,\nn
\langle {\cal O}_{abP}(x) {\cal
  O}^\dagger_{abP}(0)\rangle_{C1S2}&\propto& 
{\cal A}_6\cos(2k_{ab}x)|x|^{-K_{2,c}-1/K_{2,c}}+{\cal A}_6|x|^{-K_{2,s}-1/K_{2,c}} + \ldots,
\label{C1S2ops}
\eea
where ${\cal A}_j$ are unknown amplitudes, $\xi_b$ is the bonding
charge boson correlation length and $K_{2,c}$ ($K_{2,s}$) is the
Luttinger parameter for antibonding charge (spin) sector. 

In this section we present DMRG results for the $64\times2$ extended-Hubbard ladder 
with $t=2t_\perp=1$, $U=8V_\perp=8V_\parallel=4$ and applied external potential
of period $4k_b=\pi$ and amplitude $W_1=W_2=1.5$. The chemical potential
$\mu_-$ has been adjusted so that the total electron number is $N=90$ with
the bonding band at quarter-filling. Up to $m=1200$ density matrix states were
kept in the simulations, leading to truncation errors of $\sim10^{-6}$. This corresponds
to a relative error in the ground state energy per site of $\epsilon\approx10^{-4}$.

The presence of a charge gap in the bonding sector is confirmed by the
examination of the Green's functions in the bonding ($G_b(n)$) and
antibonding ($G_{ab}(n))$ bands. The RG analysis suggests that $G_{ab}(n)$
decays as a power law, whereas $G_{b}(n)$ decreases with distance as
an exponential multiplied by a power law. 

\begin{figure}[ht]
\begin{tabular}{ccc}
(a)\includegraphics[width=0.46\textwidth]{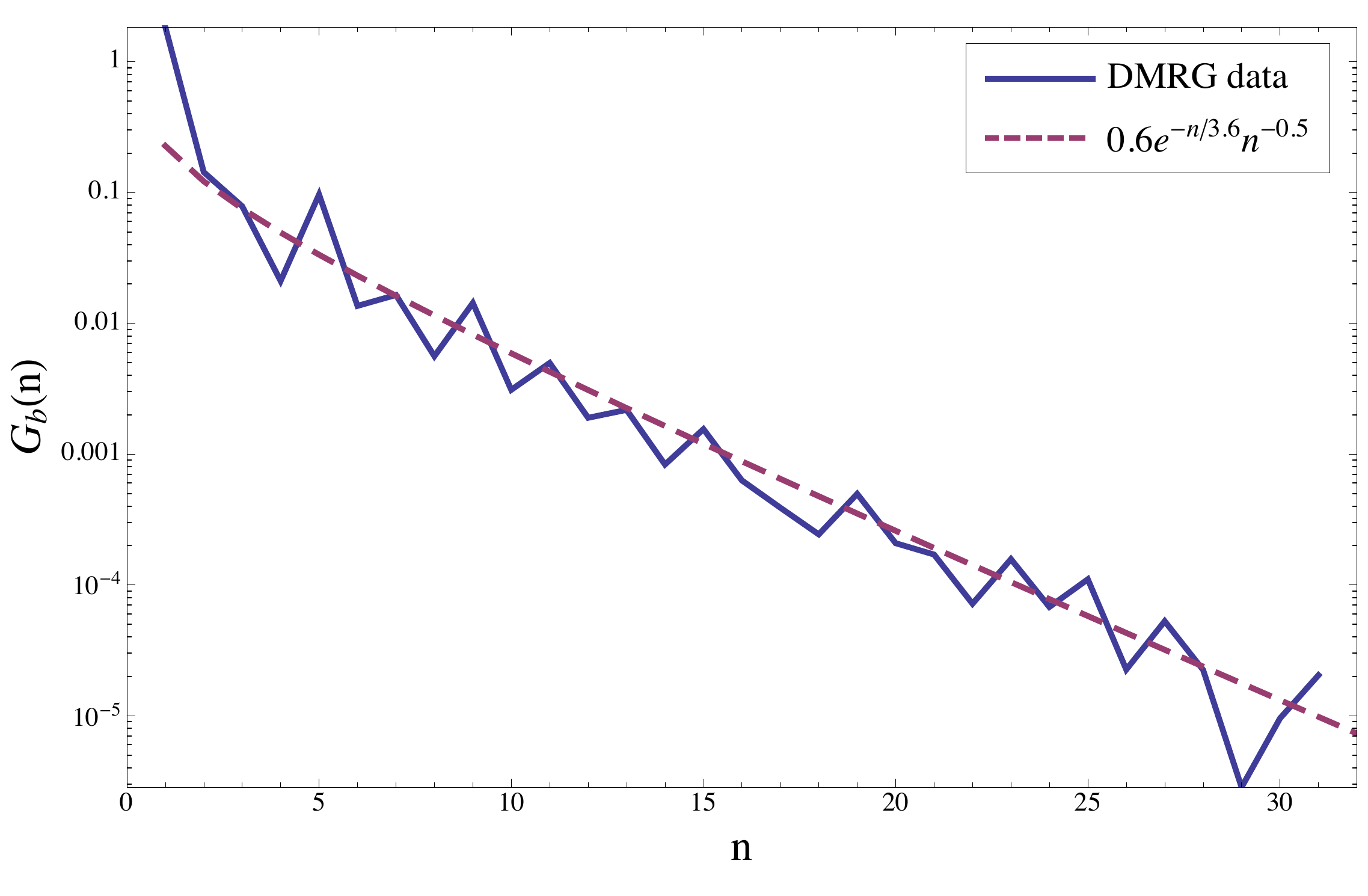}&\quad&
(b)\includegraphics[width=0.46\textwidth]{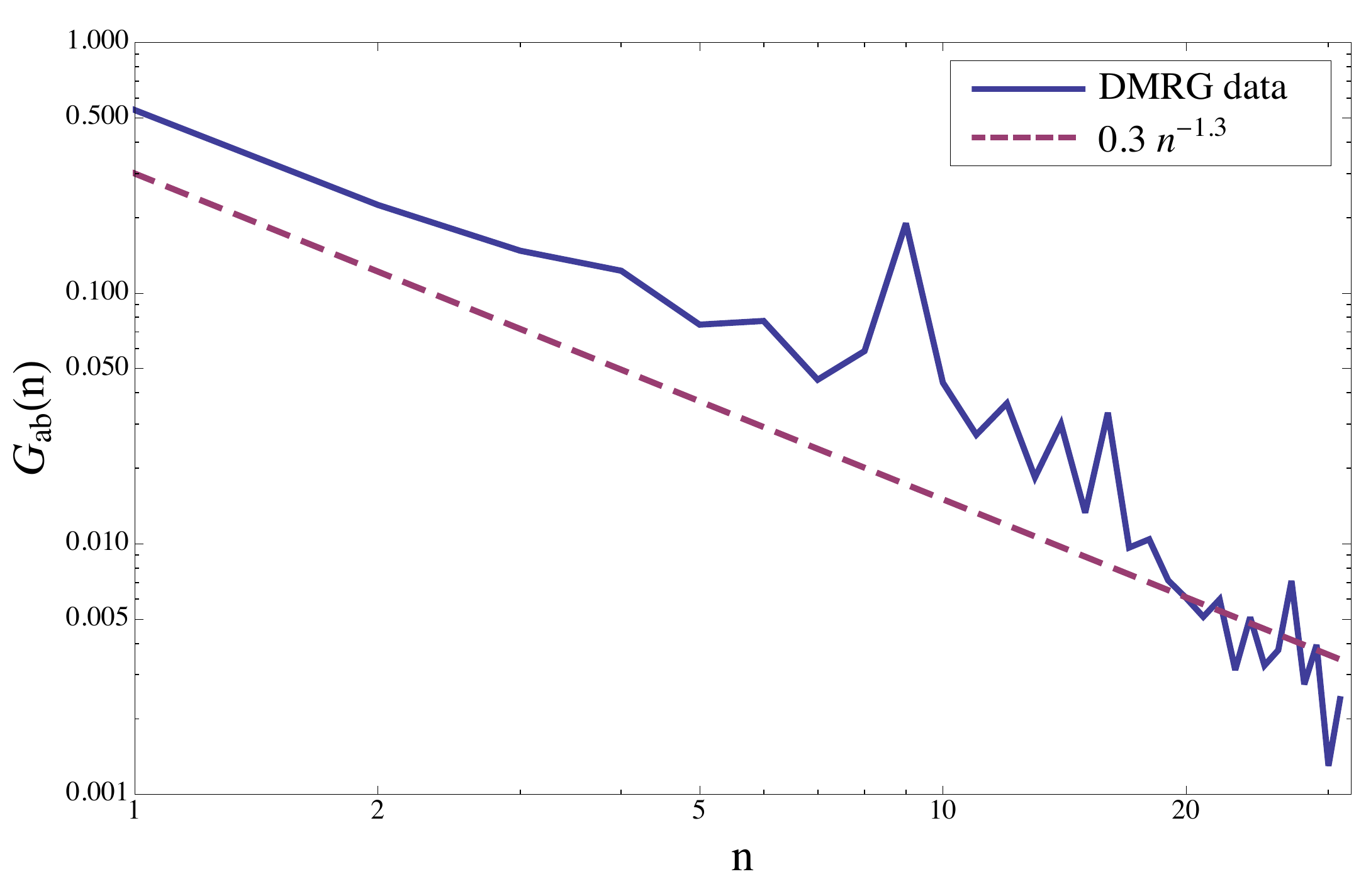}
\end{tabular}
\caption{DMRG data (solid) and fits (dashed) for 
(a) the bonding Green's function $G_{b}(n)=\langle c_{b}(32)c_{b}^\dagger(32+n)\rangle$
with $k_b$ oscillations removed (see text),
and (b) the antibonding Green's function 
$G_{ab}(n)=\langle c_{ab}(32)c_{ab}^\dagger(32+n)\rangle$
with $k_{ab}$ oscillations removed. Additional oscillations
in both cases are due to sub-leading contributions of the Green's 
functions. For both figures the $64\times2$ ladder is considered with 
$t=2t_\perp=1$, $U=8V_\perp=8V_\parallel=4$ and $W_+=1.5$.
The chemical potential $\mu_-$ was chosen such that the bonding band is quarter-filled
for the system with $N=90$ electrons. Fit functions are of the form 
predicted by the RG analysis.}
\label{fig:GFs}
\end{figure}

The bonding Green's function is shown in Fig.~\ref{fig:GFs}(a), where the leading oscillations
at $k_b$ have been removed by performing a fit to the Green's function and dividing out
the oscillatory part. So, in Fig.~\ref{fig:GFs}(a) we plot
\be
G_{b}(n) = \frac{G_{b,{\rm Full}}(n)}{\cos\left(k_b x\right)},\nonumber
\ee
where $G_{b,{\rm Full}}(n)$ is the full bonding Green's function with oscillations at $k_b$. 
The leading oscillation has been removed in order to elucidate
the long-distance behaviour of the Green's function. In this case the asymptotic behaviour
is well described by an exponential multiplied by a power law, as predicted by
the RG analysis. We perform a similar procedure for the antibonding Green's function
in Fig.~\ref{fig:GFs}(b), where the leading oscillations occur at $k_{ab}$. The power law
decay of the antibonding Green's function is in agreement with the RG analysis.
The form of both Green's functions is consistent with the expectations of 
the $C1S2$ phase, with a single massive charge boson in the bonding sector 
of the theory.

\begin{figure}[ht]
\begin{tabular}{ccc}
(a)\includegraphics[width=0.46\textwidth]{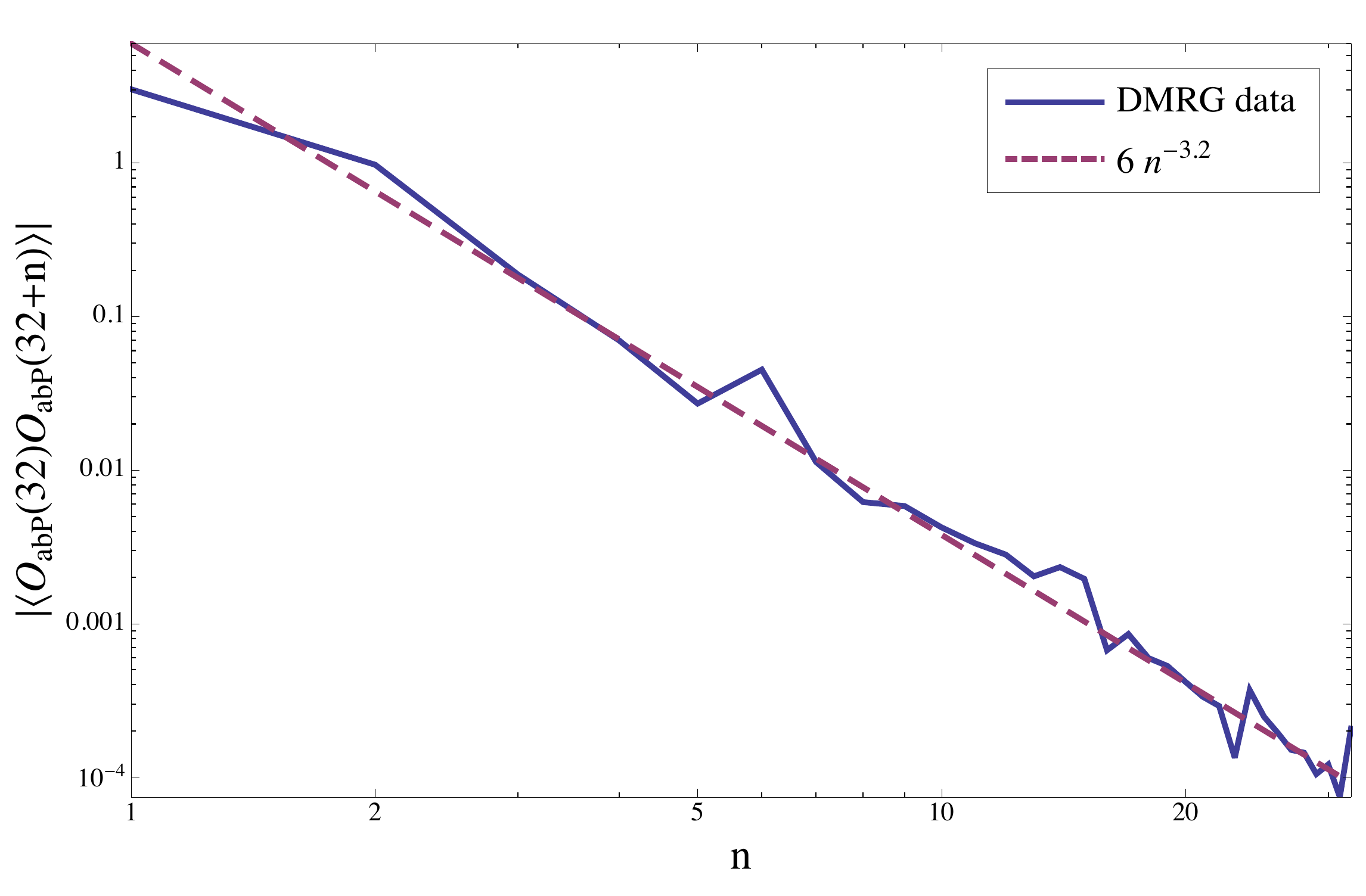} &\quad&
(b)\includegraphics[width=0.46\textwidth]{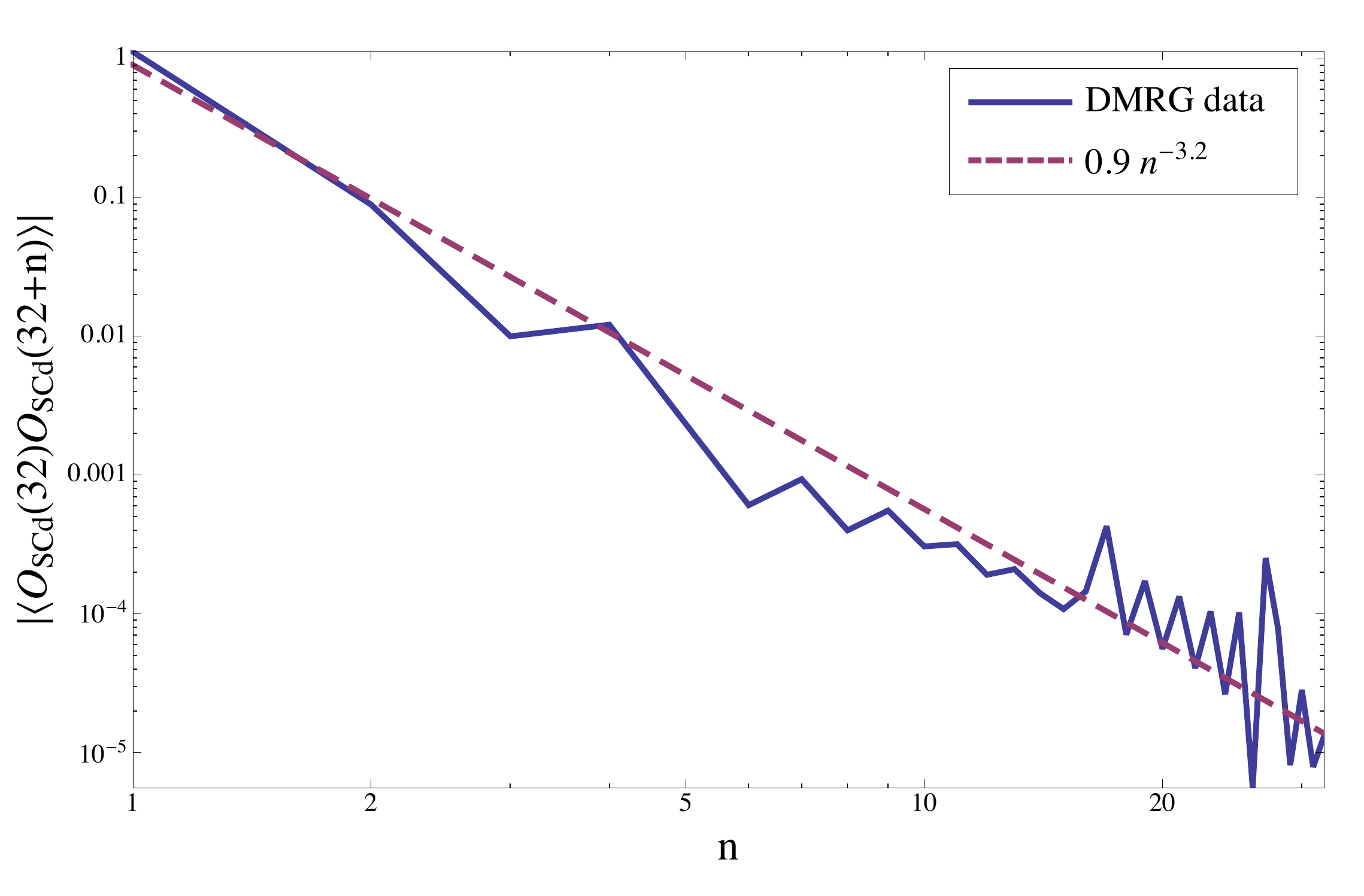}
\end{tabular}
\caption{DMRG data (solid) and fit functions (dashed) for
(a) the two-point function of the antibonding pairing order parameter
${\cal O}_{abP}$ with the $2k_{ab}$ oscillations removed (see text), 
and (b) the two-point function of the superconducting d-wave 
order parameter ${\cal O}_{SCd}$ with $2k_{ab}$ oscillations removed
 on the $64\times2$ ladder with $t=2t_\perp=1$, 
$U=8V_\perp=8V_\parallel=4$ and $W_+=1.5$. The chemical potential has been
adjusted so that $N=90$ coincides with a quarter-filled bonding band.
Fit functions are of the form given in Eqs.~\fr{C1S2ops}.}
\label{fig:C1S2tpf}
\end{figure}

Having established the presence of a charge gap in the bonding sector,
we now consider the two-point functions of the order parameters~\fr{C1S2ops}, 
shown in Fig.~\ref{fig:C1S2tpf}. As with our analysis of the Green's function,
the two-point functions of the antibonding pairing order parameter and the superconducting 
d-wave order parameter, shown in Fig.~\ref{fig:C1S2tpf}(a) and Fig.~\ref{fig:C1S2tpf}(b) respectively,
have had the leading order $2k_{ab}$ oscillation removed.
Both two-point functions show power law decay with the same exponent,
giving an approximate value for the Luttinger parameter in the antibonding charge 
sector $K_{2,c}\approx0.35$.  

\begin{figure}[ht]
\includegraphics[width=0.52\textwidth]{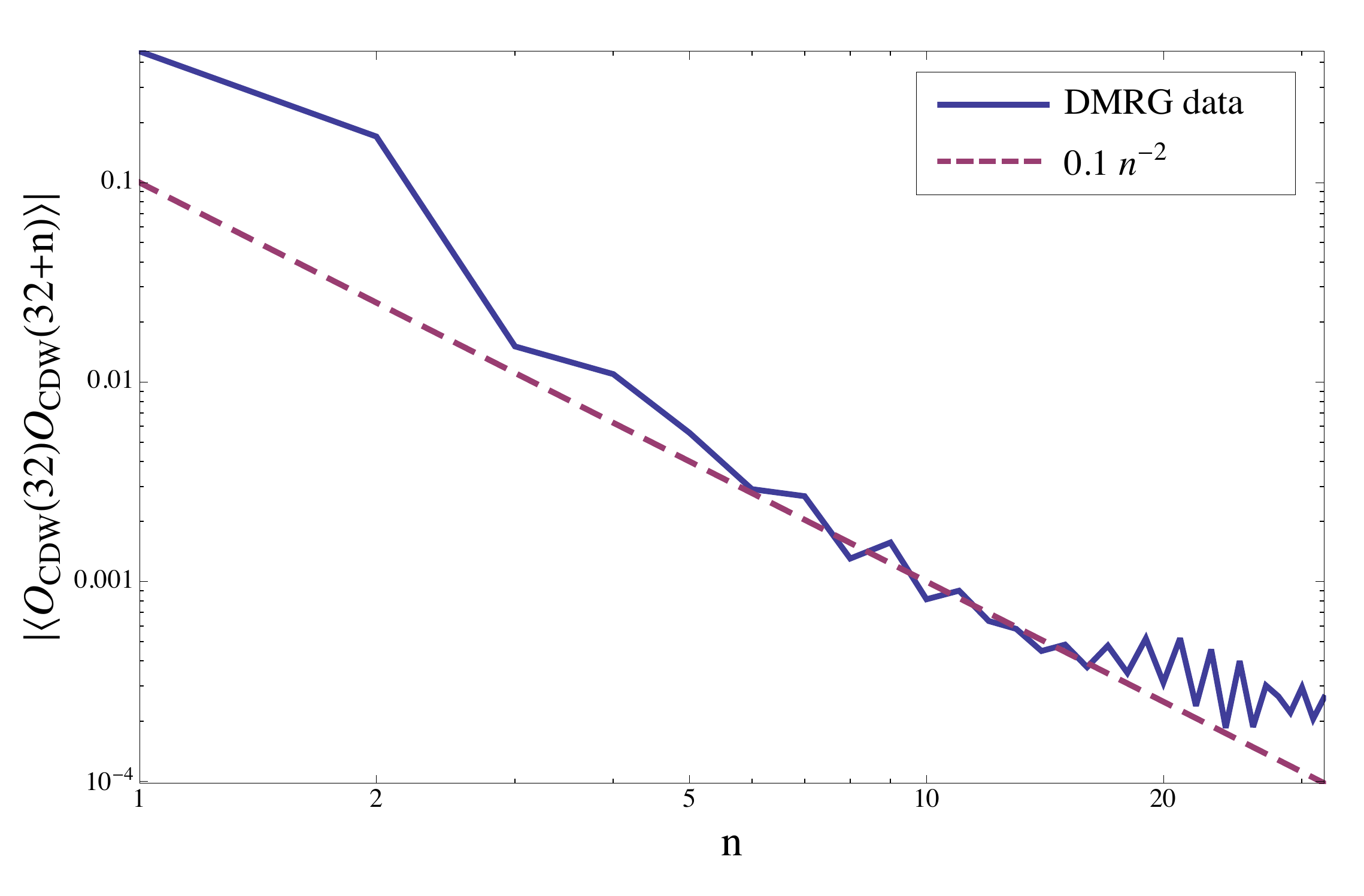}
\caption{DMRG data (solid) and $x^{-2}$ guide (dashed) for
the two-point function of the charge density wave order
parameter ${\cal O}_{CDW}$ on the $64\times2$ ladder with $t=2t_\perp=1$, 
$U=8V_\perp=8V_\parallel=4$ and $W_+=1.5$. The chemical potential has been
adjusted so that $N=90$ coincides with a quarter-filled bonding band. Sub-$x^{-2}$
decay is observed with oscillations at $2k_{ab}$ at large distances.}
\label{fig:C1S2cdw}
\end{figure}

Figure~\ref{fig:C1S2cdw} shows the two-point function of the 
charge density wave (CDW) order parameter. At intermediate distances this
is well described by $x^{-2}$ decay, whilst for large distances it decays at 
slower-than-$x^{-2}$ and oscillates with wavenumber $2k_{ab}$, 
as predicted from the bosonization analysis~\fr{C1S2ops}. Sub-leading
$4k_b$ contributions are also observed. The long-distance
decay suggests that the Luttinger parameter in the antibonding spin sector is
$K_{2,s}=1$, as expected from $SU(2)$ symmetry.  The dominant
correlations for considered parameters are of the charge density wave type.

\subsubsection{$C1S0$ $4k_b$ Mott Insulator Phase}

As has been discussed in Section~\ref{C1S0}, in order 
for the $C1S0$ $4k_b$ Mott insulating phase to occur, it
is necessary for the interchain exchange interaction to
be \emph{antiferromagnetic} after the initial renormalization
group procedure. This can always be achieved provided
the interchain density-density interaction coupling is large 
$V_\perp>U$, such that for the initial conditions the exchange 
interaction is antiferromagnetic and remains so under the 
renormalization group procedure.

At the $C1S0$ fixed point, the $4k_b$ Mott insulator phase
is characterized by the following asymptotic forms of the two-point functions
 \bea
\langle {\cal O}_{CDW}(x)\ {\cal O}^\dagger_{CDW}(0)\rangle
&\propto& {\cal A}_8x^{-2}+{\cal A}_9\cos\big(2(k_{b}+k_{ab})x\big)|x|^{-K_{2,c}}+\ldots,\nn 
\langle {\cal O}_{bCDW}(x)\ {\cal  O}^\dagger_{bCDW}(0)\rangle
&\propto& e^{-|x|/\xi_b}~~({\rm  at~2}k_b)+\ldots,\nn 
\langle {\cal O}_{SCd}(x)\ {\cal  O}^\dagger_{SCd}(0)\rangle
&\propto&
\cos(2k_{ab}x)|x|^{-K_{2,c}-1/K_{2,c}}
+\ldots\ ,\nn
\langle {\cal O}_{abP}(x)\ {\cal   O}^\dagger_{abP}(0)\rangle&\propto&
{\cal A}_{10}\cos(2k_{ab}x)|x|^{-K_{2,c}-1/K_{2,c}} +  {\cal A}_{11} \cos\big(2k_b x\big)|x|^{-1/K_{2,c}}+\ldots
\ , \label{4kbops}
\eea
where ${\cal A}_d$ are unknown amplitudes. 

We present results for the Hamiltonian~\fr{Hpi} on 
on the $96\times2$ ladder with $t=2t_\perp=1$, $V_\perp=5$, 
$V_\parallel=0$ and $U=4$. The chemical potential $\mu_-$ is 
used to set the total number of electrons to $N=88$ whilst 
maintaining the bonding band at quarter-filling. A periodic 
potential with period $4k_b=\pi$ and amplitude $W_+=1$ is
applied to the bonding band. Up to $m=2000$ density 
matrix states were kept in the calculations, giving truncation
errors of $\sim10^{-7}$. The increased number of states in the procedure
results in a relative error for the ground state energy per site of
$\epsilon\approx2\times10^{-5}$.

\begin{figure}[ht]
\begin{tabular}{ccc}
(a)\includegraphics[width=0.45\textwidth]{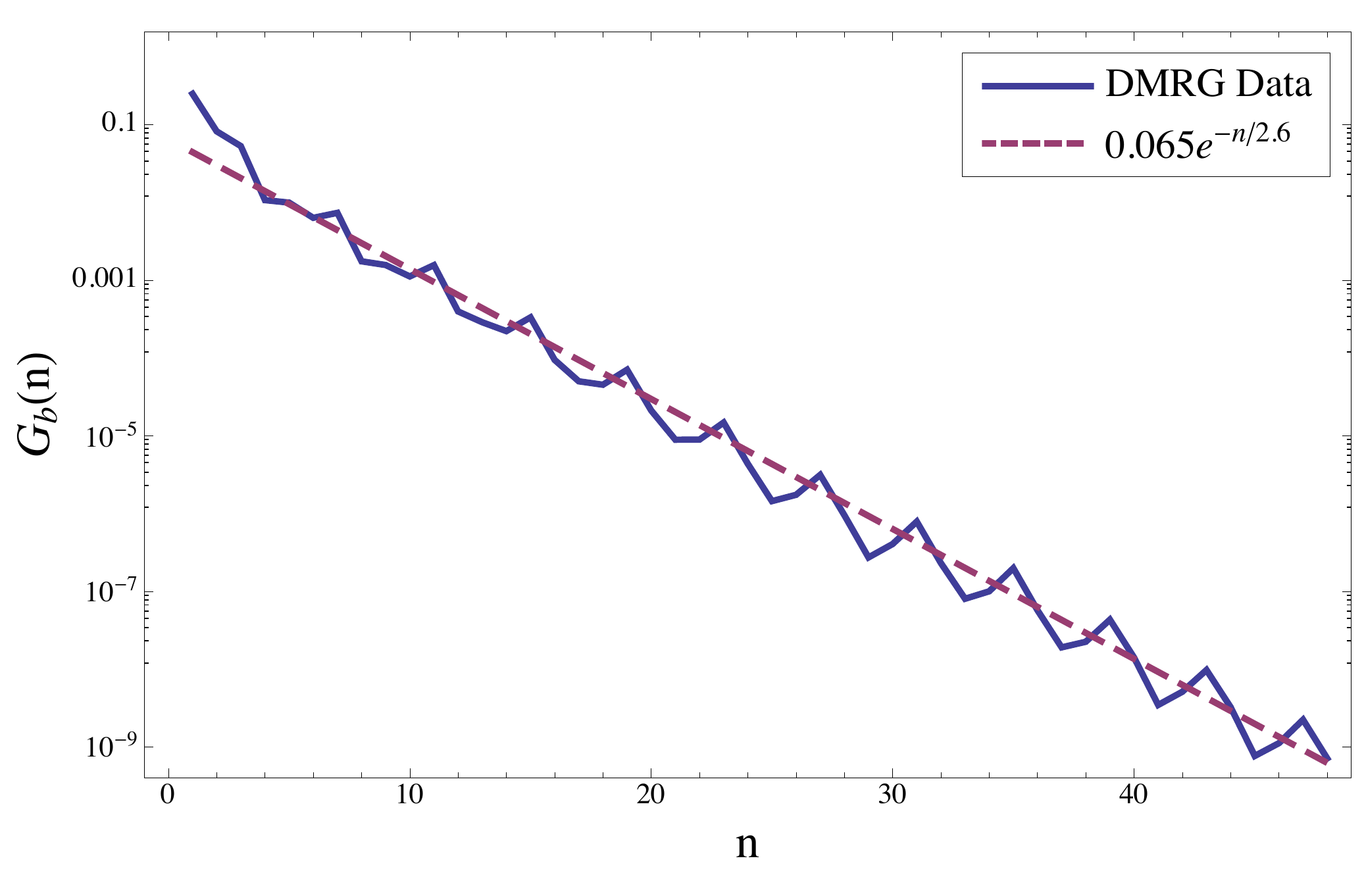} &\quad&
(b)\includegraphics[width=0.45\textwidth]{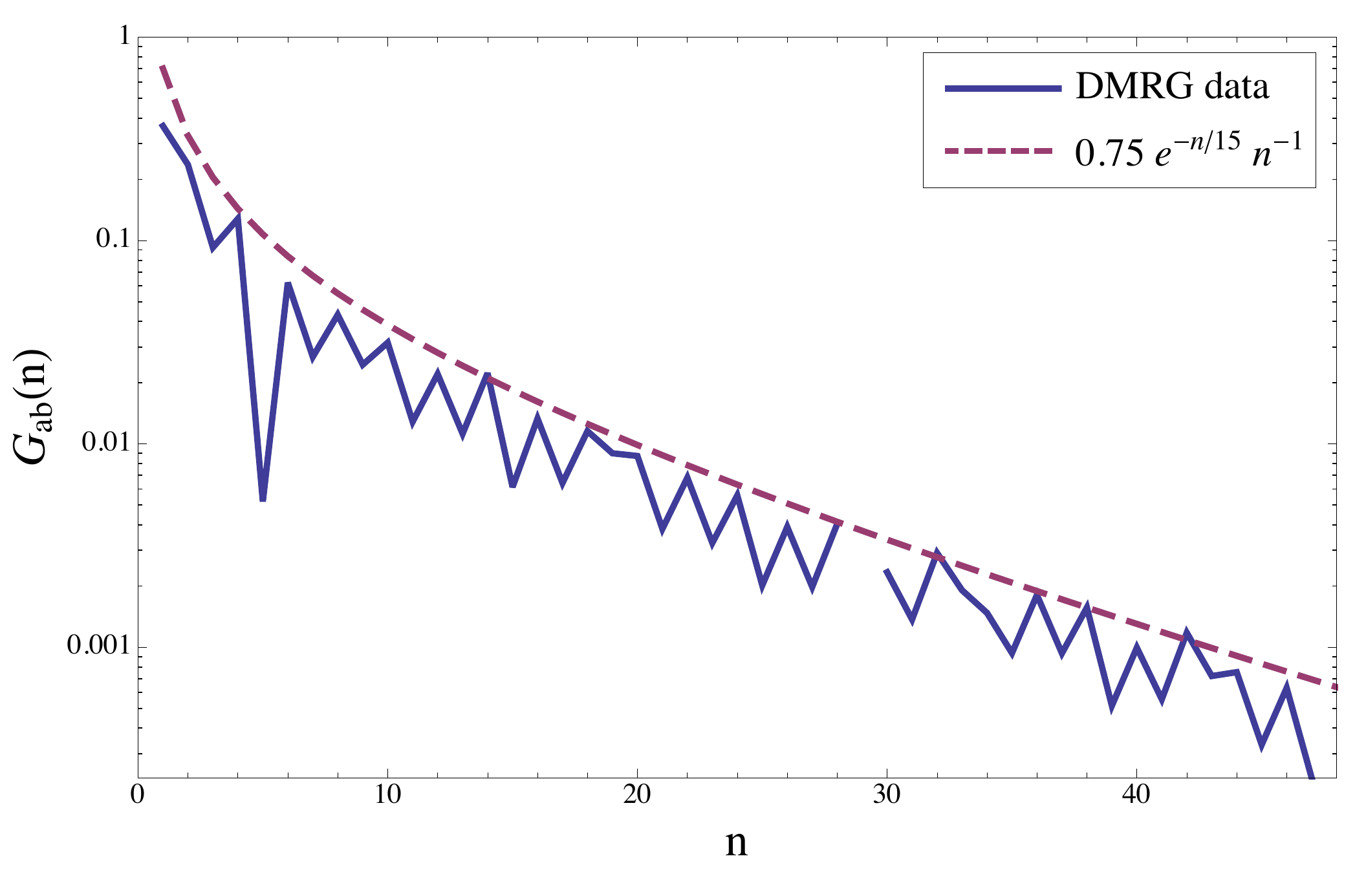}
\end{tabular}
\caption{DMRG data (solid) and fit function (dashed) for
(a) the bonding Green's function 
$G_{b}(n) = \langle c_{b}(48)c_{b}^\dagger(48+n)\rangle$,
and (b) the antibonding Green's function
$G_{ab}(n) = \langle c_{ab}(48)c_{ab}^\dagger(48+n)\rangle$,
with the $k_{ab}$ oscillation removed (see text), for the 
$96\times2$ ladder with $t=2t_\perp=1$, $U=4$,
$V_\parallel=0$, $V_y=5$, $W_+=1$ and $N=88$ electrons. The chemical
potential $\mu_-$ was chosen such that this corresponds to 
a quarter-filled bonding band. Oscillations in both plots
are from other contributions to the Green's function.
The break in data of (b) at $n=29$ is a remnant of removing the $k_{ab}$ oscillations.
 }
\label{fig:C1S0_GFs}
\end{figure}

The presence of a spin gap in both bands and a 
charge gap in the bonding band is inferred from the
forms of the two-point functions~\fr{4kbops} and the 
Green's functions shown in Fig.~\ref{fig:C1S0_GFs}.
The RG analysis predicts that the bonding Green's function
should decay exponentially, whilst the antibonding Green's function
should decay as an exponential multiplied by a power law. 
In Fig.~\ref{fig:C1S0_GFs}(a) the bonding Green's function
($G_{b}(n)$) is shown with an exponential fit and is well described 
by exponential decay, implying both spin and charge gaps in the bonding sector.
Figure~\ref{fig:C1S0_GFs}(b) shows the antibonding Green's function with
with the leading oscillation at wavevector $k_{ab}$ removed in order 
to more clearly show the exponential multiplied by power law fit, as predicted
by the RG analysis.  
The break in the plot of $G_{ab}(n)$ close to $n=28$ is a result of removing 
the oscillation; for this point the fit and $G_{ab}(n)$ differ in sign whilst both 
magnitudes are close to zero. The fit gives an approximate 
value for the Luttinger parameter in the antibonding charge sector 
$K_{2,c}\approx0.27$. 

\begin{figure}[ht]
\begin{tabular}{ccc}
(a)\includegraphics[width=0.46\textwidth]{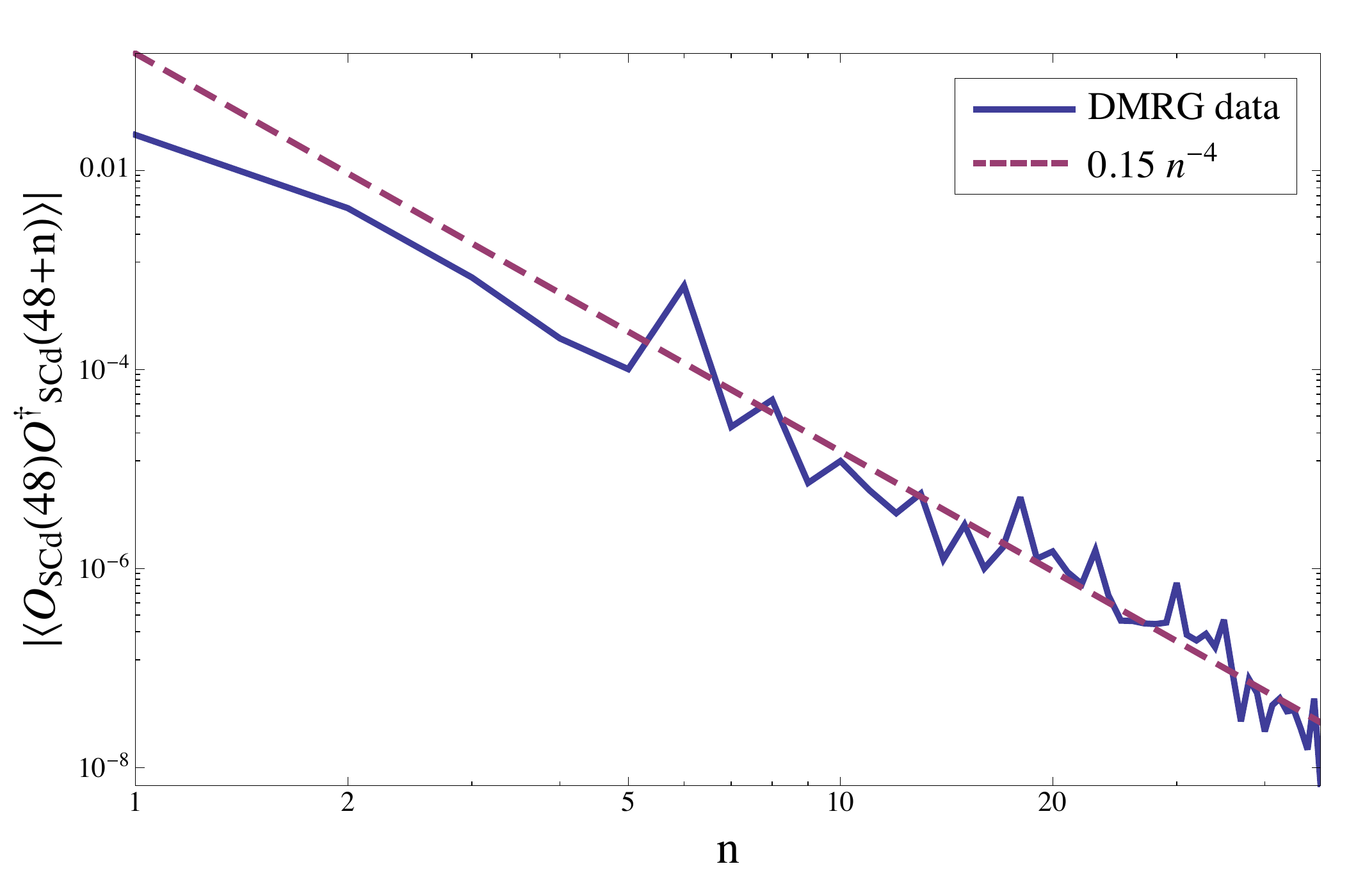} &\quad&
(b)\includegraphics[width=0.45\textwidth]{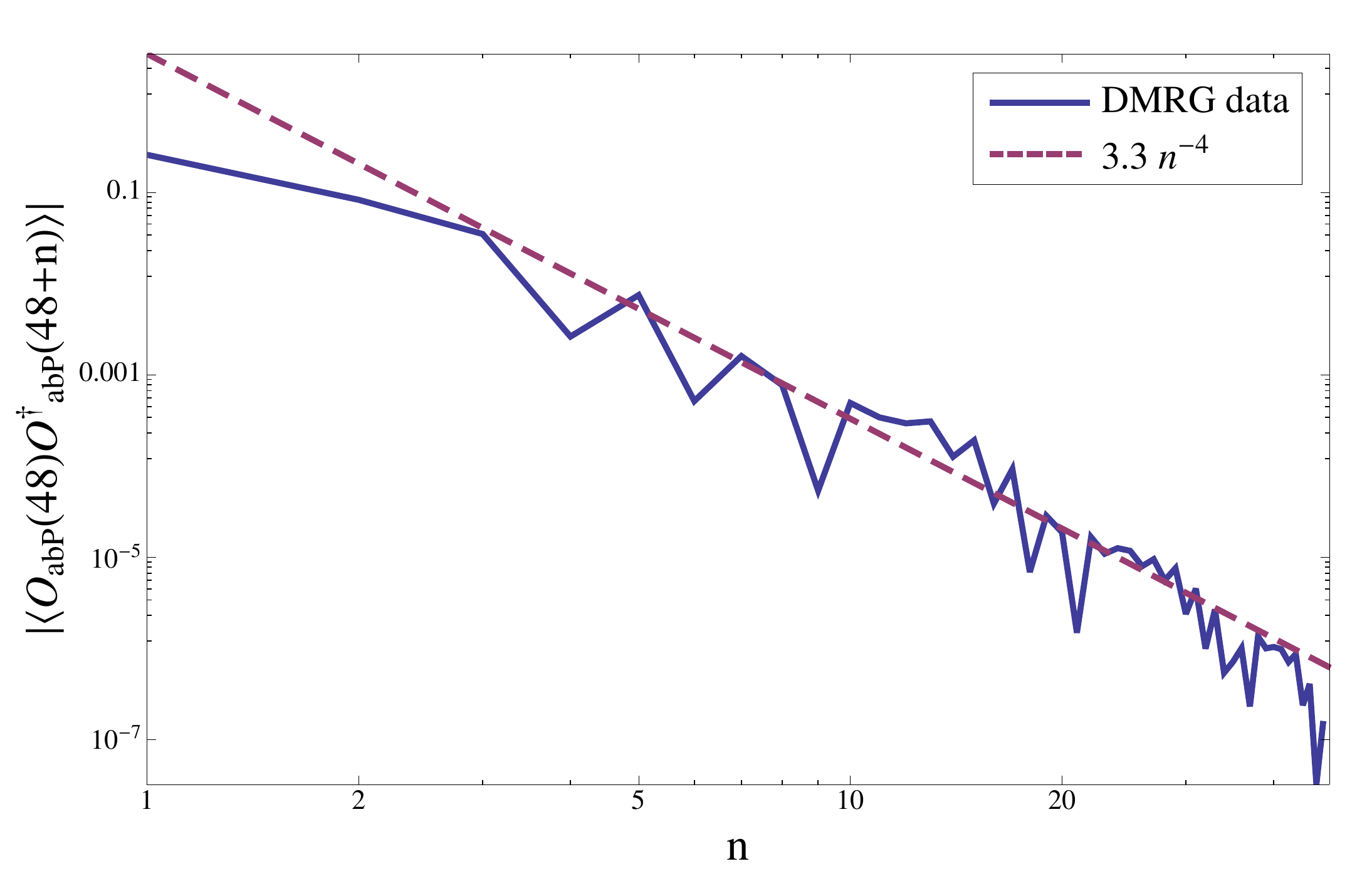}
\end{tabular}
\caption{DMRG data (solid) and fit functions (dashed)
for (a) the absolute value of the two-point function
of the superconducting d-wave order parameter ${\cal O}_{SCd}$
with $2k_{ab}$ oscillations removed (see text),
and (b) the absolute value of the two-point function of
the antibonding pairing order parameter ${\cal O}_{abP}$
on the $96\times2$ ladder with $t=2t_\perp=1$, $U=4$,
$V_\parallel=0$, $V_y=5$, $W_+=1$ and $N=88$ electrons. The chemical
potential $\mu_-$ was chosen such that this corresponds to 
a quarter-filled bonding band.  
}
\label{fig:C1S0_tpfs}
\end{figure}

With both Green's functions being consistent with the $C1S0$ phase,
the two-point functions of the order parameters in Eqs.~\fr{4kbops}
are now considered. The two-point functions for the SCd order 
parameter and the abP order parameter are presented in 
Fig.~\ref{fig:C1S0_tpfs}(a) and Fig.~\ref{fig:C1S0_tpfs}(b) respectively.
In both cases the leading oscillation at frequency $2k_{ab}$ has been
removed in order to elucidate the form of the decay, which in both 
cases is well described by a power law with an exponent consistent 
with $K_{2,c}\approx0.27$. The absence of power law decay with
exponent $1/K_{2,c}$ for the antibonding pairing order parameter
is not inconsistent with being in the $C1S0$ phase, as the amplitude
${\cal A}_{11}$ is interaction-dependent and may be much smaller than
the amplitude ${\cal A}_{10}$ of the sub-leading decay, in which case
at short-distances the sub-leading decay would dominate. 

\begin{figure}[ht]
\includegraphics[width=0.46\textwidth]{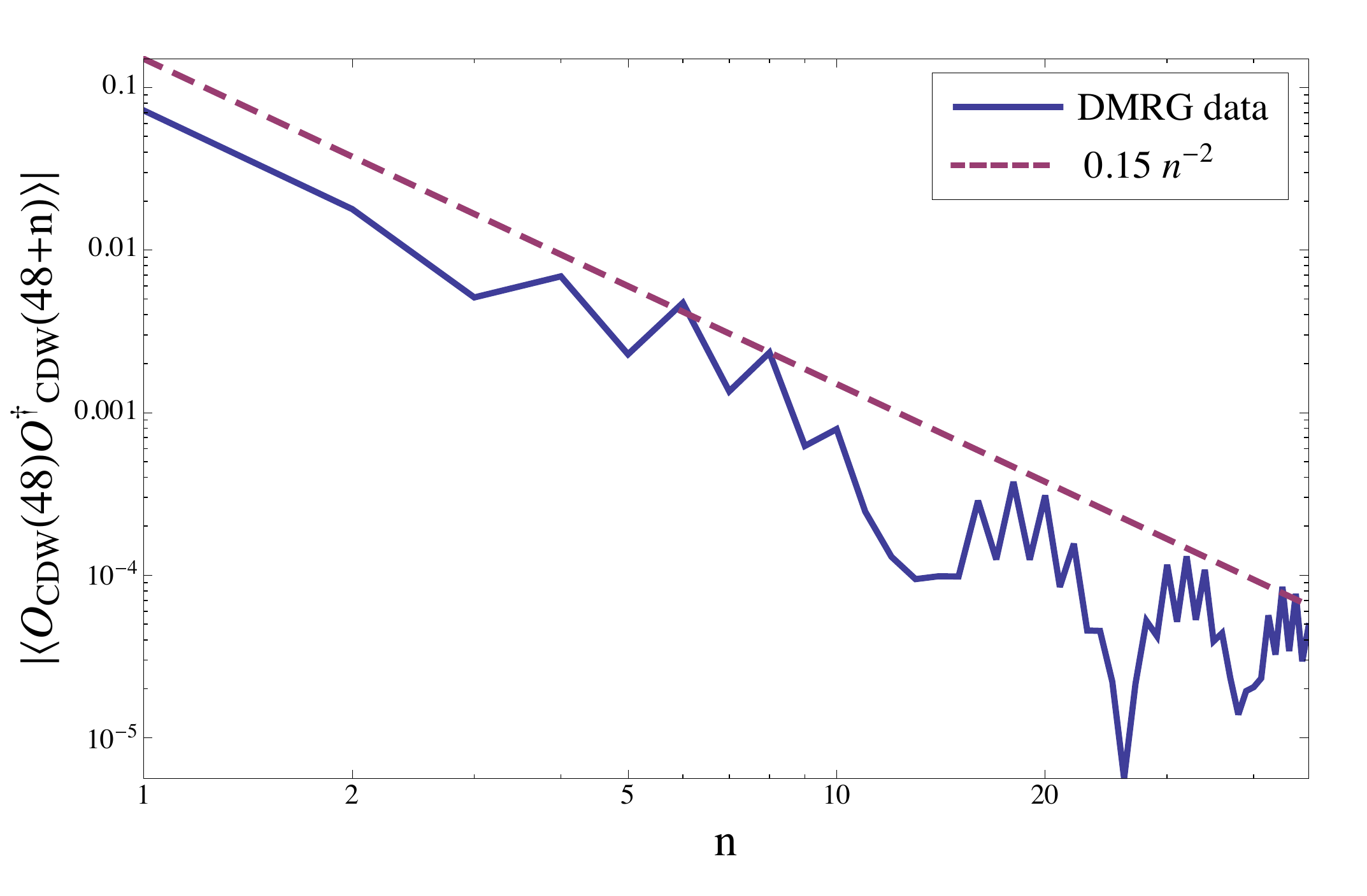}
\caption{The two-point function of charge density wave
order parameter ${\cal O}_{CDW}$ (solid) and $x^{-2}$ power law (dashed) for
the $96\times2$ ladder with $t=2t_\perp=1$, $U=4$,
$V_\perp=5$ and $W_+=1$. A chemical potential applied to the 
antibonding band is used to set the electron number
to $N=88$ with the bonding band quarter-filled. }
\label{fig:C1S0_cdw}
\end{figure}

The two-point function of the charge density wave order parameter is 
shown in Fig.~\ref{fig:C1S0_cdw}. At long distances there are large
wavelength oscillations with wavevector $2k_b+2k_{ab}$ decaying 
at sub-$x^{-2}$, consistent with the bosonization predictions for the 
$C1S0$ phase~\fr{4kbops}. The exact form of the decay of the 
$2k_b+2k_{ab}$ oscillations cannot be accurately extracted in the
$L=96$ system, due to the large spin correlation length and 
the amplitudes ${\cal A}_8$ and ${\cal A}_9$ being unknown. 

The two-point function of the bonding charge density wave order
parameter can also be calculated, however information is not 
easily extracted from this two-point function due to the long spin 
correlation length and unknown interaction-induced amplitudes 
of $4k_F$ components of the bonding charge density operator, 
which are similar in form to those in Eqs.~\ref{4kfbosons}. 

As discussed in detail in Section~\ref{C1S0}, there are two
possibilities for the dominant correlation in the $4k_b$ Mott insulator, 
depending upon $K_{2,c}$. For the presented case, $K_{2,c}<1$
and the dominant correlations are of charge density wave type,
arising from the interaction-induced $2k_b+2k_{ab}$ component 
of the charge density.

\section{Conclusions}
\label{sec:Conclusions}
In this work we have established a mechanism for finite wavevector
pairing in doped fermionic ladders with equivalent legs. This
mechanism is driven by Umklapp scattering processes, which occur
either at special band fillings as a result of electron electron
interactions, see also \cite{FradkinArx},  or are induced by
``externally'' applied periodic potentials. The latter can arise via
charge-density wave formation driven by the (three-dimensional)
long-ranged Coulomb interaction in real crystal structures. We have
applied renormalization group methods in the low-energy limit of the
lattice model \fr{Hamiltonian} for (i) weak interactions ("band
representation'') and (ii) arbitrary interaction strength but very
small tunneling along the rung direction (``chain
representation''). In both cases we have found that the theory
describing the strong coupling fixed point is the same as the low
energy description of the so-called Kondo-Heisenberg Model
(KHM)~\cite{zachar,berg}. In the case of the $4k_b$ Mott insulator
analyzed in section \ref{sec:4kb} this fact may be anticipated on the
basis of the following arguments. The $4k_b$ Umklapp scattering
process leads to formation of a Mott gap $\Delta$ within the bonding
band. At low-energies the charge dynamics is the blocked by the Mott
gap and at energies small compared to $\Delta$ one is left with spin
degrees of freedom, that can be thought of in terms of an effective
spin-1/2 Heisenberg chain. The antibonding degrees of freedom remain
gapless, and at low energies compared to $\Delta$ the most important
interaction with the bonding degrees of freedom is then through an
effective spin exchange interaction. The resulting picture is an
effective KHM, where the spin-1/2 chain corresponds to the bonding
band and the role of the interacting 1D electron gas is played by the
antibonding band. The low energy limit is crucial for these
considerations to hold, because in the lattice model \fr{Hamiltonian}
electron number in the bonding band is not conserved. 

Another important difference between the effective KHM
that emerges as the low-energy description of the ladder and the
lattice KHM model considered in \cite{zachar,berg} is that the effective
exchange interaction between the bonding and antibonding bands is not
\emph{a priori} antiferromagnetic. In the case of weakly interacting
Hubbard chains it is in fact \emph{ferromagnetic}, which results in a
C1S2 phase as the exchange interaction is marginally
irrelevant. On the other hand, we found that extended density-density
interactions (we explicitly consider repulsive nearest-neighbor
interactions) can cause this exchange interaction to become
antiferromagnetic. In this case the low-energy sector of the 
theory is a C1S0 phase, where the remaining gapless degree of freedom
describes the antibonding charge sector and is characterized by its
Luttinger parameter $K_{2,c}$. The dominant correlations are then either
of superconducting PDW (if $K_{2,c}>1$) or CDW (if $K_{2,c}<1$) type.

The activation of the Umklapp scattering process at $3k_b+k_{ab}$ results in 
a similar low-energy description, although here the remaining massless degree
of freedom $\tilde\Phi_2$ is significantly more complicated: it
is a combination of the symmetric charge boson $\Phi_c$ and the U(1)
doublet Majorana fermions $\xi^{1,2}$, which are themselves comprised
of the SU(2) singlet Majorana fermion from the antisymmetric spin
sector and a Majorana fermion from the antisymmetric charge
sector. The composite nature of this gapless degree of freedom makes
the analysis of ground state correlations difficult and we leave this
issue to future studies.

\acknowledgments
We thank E. Fradkin, A.A. Nersesyan and D.A. Tennant for valuable
discussions. NJR and FHLE are supported by the EPSRC under grant
EP/I032487/1. AMT thanks the Rudolf Peierls Centre for Theoretical
Physics for hospitality and acknowledges support from the Center for
Emergent Superconductivity, an Energy Frontier Research Center funded
by the US Department of Energy, Office of Science, Office of Basic
Energy Sciences. We are grateful to the Aspen Center for Physics and
NSF grant 1066293 for hospitality and support.

\appendix
\section{The Charge Density Operator}
\label{app:CD}
At commensurate fillings, or by applying an appropriate external
periodic potential, Umklapp scattering processes can be activated in
the doped ladder. In this case, oscillatory components of the charge density which
are usually suppressed away from commensurate fillings now feature in the 
Hamiltonian. In this appendix we consider the $2k_F$ and $4k_F$
harmonics of the charge density operator in the ``band'' and ``chain'' 
representations in turn.

\subsection{$2k_F$ Components of the Charge Density}
\label{app:2kF}
We first consider the $2k_F$ harmonics in the ``band'' representation.
The number operators on each leg of the ladder can be expressed in
terms of the bonding/antibonding fermions introduced in~\fr{babvar}
as
\bea
n_{1,j,\alpha}+n_{2,j,\alpha}&=&c^\dagger_{b,j,\alpha}c_{b,j,\alpha}
+c^\dagger_{ab,j,\alpha}c_{ab,j,\alpha}\ ,\nn
n_{1,j,\alpha}-n_{2,j,\alpha}&=&c^\dagger_{b,j,\alpha}c_{ab,j,\alpha}
+c^\dagger_{ab,j,\alpha}c_{b,j,\alpha}\ .
\eea
Linearizing about the Fermi surface and taking the continuum limit
as in~\fr{continuum}, we obtain the following decompositions
\bea
n_{1,j,\alpha}+ n_{2,j,\alpha}&\sim&a_0
\rho^{(+)}_{0,\alpha}(x)+a_0\left[\rho^{(+)}_{2k_b,\alpha}(x)
\ e^{2ik_bx}
+\rho^{(+)}_{2k_{ab},\alpha}(x)
\ e^{2ik_bx}
+{\rm h.c.}\right]+\ldots\ ,\nn
n_{1,j,\alpha}- n_{2,j,\alpha}&\sim&a_0\left[
\rho^{(-)}_{k_b+k_{ab},\alpha}(x)
\ e^{i(k_b+k_{ab})x}
+{\rm h.c.}\right]+\ldots,
\eea
where
\bea
\rho^{(+)}_{0,\alpha}(x) &=& R^\dagger_{b,\alpha}R_{b,\alpha}
+L^\dagger_{b,\alpha}L_{b,\alpha} +R^\dagger_{ab,\alpha}R_{ab,\alpha}+
L^\dagger_{ab,\alpha}L_{ab,\alpha} \ ,\nn
\rho^{(+)}_{2k_{b},\alpha}(x)&=&
L^\dagger_{b,\alpha}R_{b,\alpha}\ ,\
\rho^{(+)}_{2k_{ab},\alpha}(x)=
L^\dagger_{ab,\alpha}R_{ab,\alpha}\ ,\nn
\rho^{(-)}_{k_{b}+k_{ab},\alpha}(x)&=&
L^\dagger_{ab,\alpha}R_{b,\alpha}
+L^\dagger_{b,\alpha}R_{ab,\alpha}\ .
\eea
We note that $\rho_{0,\alpha}(x)$, $\rho_{2k_{b},\alpha}(x)$ 
and $\rho_{2k_{ab},\alpha}(x)$ are
even under interchange of legs 1 and 2 of the ladder, while
$\rho_{k_{b}+k_{ab},\alpha}(x)$ is odd. The components can then be
bosonized following~\cite{GNT} and
(\ref{bosonization1}-\ref{spinchargefields}). This leads to the 
following expressions for components of the charge density operator
\bea
\rho^{(+)}_{0}(x) &=& \sum_\alpha\rho^{(+)}_{0,\alpha}(x)
\sim \frac{1}{\sqrt{2\pi}}\partial_x \Phi_{+c}\ , \nn
\rho^{(+)}_{2k_b}(x)&=&\sum_\alpha\rho^{(+)}_{2k_b,\alpha}(x)
\sim-2i\ e^{i\sqrt{\pi}(\Phi_{+,c}+\Phi_{-,c})}
\cos\Big[\sqrt{\pi}(\Phi_{+,s}+\Phi_{-,s})\Big]\ ,\nn
\rho^{(+)}_{2k_{ab}}(x)&=&\sum_\alpha\rho^{(+)}_{2k_{ab},\alpha}(x)
\sim-2i\ e^{i\sqrt{\pi}(\Phi_{+,c}-\Phi_{-,c})}
\cos\Big[\sqrt{\pi}(\Phi_{+,s}-\Phi_{-,s})\Big]\ ,\nn
\rho^{(-)}_{k_b+k_{ab}}(x)
&\sim&-2i\ e^{i\sqrt{\pi}\Phi_{+,c}}
\Big[e^{-i\sqrt{\pi}\Theta_{-,c}}\cos\left(\sqrt{\pi}[\Phi_{+,s}
-\Theta_{-,s}]\right)
-e^{i\sqrt{\pi}\Theta_{-,c}}\cos\left(\sqrt{\pi}[\Phi_{+,s}
+\Theta_{-,s}]\right)\Big],
\eea
where $\Phi_{\pm,d}=(\Phi_{1,d}\pm\Phi_{2,d})/\sqrt{2}$ and 
$\Theta_{\pm,d}=(\Theta_{1,d}\pm\Theta_{2,d})/\sqrt{2}$ for $d=c,s$.
In the final term we have used that $\kappa_{b,\uparrow}
\kappa_{ab,\uparrow}\equiv\kappa_{b,\downarrow}\kappa_{ab,\downarrow}$ and
$(\kappa_{b,\sigma}\kappa_{ab,\sigma})^2=-1$. Here we note
that the $2k_F$ response of the charge density in spin gapped
phases is blocked as each term features a spin boson.

Having moved to a new basis of bosons, the $\pm$ bosons, 
we can consider refermionizing the spin bosons and the antisymmetric
charge bosons using the using the identities~\cite{GNT}
\bea
e^{i\sqrt{\pi}\Phi_{+,s}}&\sim&\mu_1\mu_2+i\sigma_1\sigma_2\ ,\qquad
e^{i\sqrt{\pi}\Theta_{+,s}}\sim\sigma_1\mu_2+i\mu_1\sigma_2\ ,\nn
e^{i\sqrt{\pi}\Phi_{-,s}}&\sim&\mu_3\mu_4+i\sigma_3\sigma_4\ ,\qquad
e^{i\sqrt{\pi}\Theta_{-,s}}\sim\sigma_3\mu_4+i\mu_3\sigma_4\ ,\nn
e^{i\sqrt{\pi}\Phi_{-,c}}&\sim&\mu_5\mu_6+i\sigma_5\sigma_6\ ,\qquad
e^{i\sqrt{\pi}\Theta_{-,c}}\sim\sigma_5\mu_6+i\mu_5\sigma_6\ ,
\eea
where $\mu_i$ and $\sigma_i$ are Majorana fermions.
Then the $2k_F$ components of the charge density operator can be
expressed in terms of the Majorana fermions as
\bea
\rho^{(+)}_{2k_{b}}(x)&\propto& -2i\ e^{i\sqrt{\pi}\Phi_{+,c}}
\left[\mu_5\mu_6+i\sigma_5\sigma_6\right]
\left[\mu_1\mu_2\mu_3\mu_4-\sigma_1\sigma_2\sigma_3\sigma_4\right]\ ,\nn
\rho^{(+)}_{2k_{ab}}(x)&\propto& -2i\ e^{i\sqrt{\pi}\Phi_{+,c}}
\left[\mu_5\mu_6-i\sigma_5\sigma_6\right]
\left[\mu_1\mu_2\mu_3\mu_4+\sigma_1\sigma_2\sigma_3\sigma_4\right]\
,\nn
\rho^{(-)}_{k_b+k_{ab}}(x)&\propto&-4i\ e^{i\sqrt{\pi}\Phi_{+,c}}
\left[\sigma_1\sigma_2\mu_3\sigma_4\sigma_5\mu_6-i
\mu_1\mu_2\sigma_3\mu_4\mu_5\sigma_6\right]\ .\nonumber
\eea
Similar expressions are obtained in the chain description with
leg indices substituted for band indices.
\section{$4k_{F}$ Density Components in the Band Picture}
\label{app:4kF}
To derive the $4k_F$-components of the charge density, we consider the
on-site Hubbard interaction
\be
U\sum_{m=1}^L [n_{1,m,\up}n_{1,m,\down}+n_{2,m,\up}n_{2,m,\down}]\ ,
\ee
which gives a contribution $S_{\rm int}=S_{\rm int}^{(1)}+S_{\rm int}^{(2)}$ to the action
\bea
S_{\rm int}^{(1)} &=&-\frac{U}{2}\int d^2{\bf y} \left[c^\dagger_{b,\up}c_{b,\up}
+c^\dagger_{ab,\up}c_{ab,\up}\right]({\bf y})
\left[c^\dagger_{b,\down}c_{b,\down}
+c^\dagger_{ab,\down}c_{ab,\down}\right]({\bf y})\nn
S_{\rm int}^{(2)}&=&-\frac{U}{2}\int d^2{\bf y} \left[c^\dagger_{b,\up}c_{ab,\up}
+c^\dagger_{ab,\up}c_{b,\up}\right]({\bf y})
\left[c^\dagger_{b,\down}c_{ab,\down}
+c^\dagger_{ab,\down}c_{b,\down}\right]({\bf y})\ .
\label{HubIntAct}
\eea
We then decompose the continuum fields into their high and low-energy parts, e.g.
\be
c_b(x)=c_{b,<}(x)+c_{b,>}(x)\ .
\label{fermiondecomp}
\ee
The $4k_F$ components of the charge density are then found by
taking the average
\be
-\Big\langle c^\dagger_b(x)c_b(x) S_{\rm int}\Big\rangle_>
\ee
over the high-energy degrees of freedom and keep only the $4k_F$
oscillating parts. For example, we obtain a contribution
\be
\frac{U}{2}\Big\langle c^\dagger_{b,\up,<}(x)c_{b,\up,>}(x) 
\int dyd\tau\ c^\dagger_{b,\up,>}c_{b,\up,<}
\left[c^\dagger_{b,\down,<}c_{b,\down,<}
+c^\dagger_{ab,\down,<}c_{ab,\down,<}\right]
\Big\rangle_>,
\ee
where we now use that
\be
\big\langle c_{b,\up,>}(x)\ c^\dagger_{b,\up,>}(\tau,y)\big\rangle=
G_{b,>}(-\tau,x-y)=-\int_{k>\Lambda}\frac{dk}{2\pi}\ e^{-ik(x-y)-
\epsilon_b(k)\tau}
\ee
is short ranged in $\tau$, so it becomes
\be
\frac{U}{2}c^\dagger_{b,\up,<}(x)\int d\tau dy\ G_{b,>}(-\tau,x-y)
c_{b,\up,<}(y)\left[c^\dagger_{b,\down,<}c_{b,\down,<}
+c^\dagger_{ab,\down,<}c_{ab,\down,<}\right](y).
\ee
Next we linearize about the Fermi surface, which decomposes the fermion
operators into their chiral components
\be
c_{b,\up,<}(y)\simeq R_{b,\up}e^{ik_by}+L_{b,\up}e^{-ik_by}
\ee
and then we replace the arguments of the left and right moving fermions
by $x$, which is justified as the Green's function is also short-ranged in
$x-y$. Implementation of this procedure leads to the following results for
the $4k_F$-components of the charge density
\bea
\rho^{(+)}_{4k_b}(x)&\sim&{U}G_{b,>}(3k_b)\sum_{\alpha=\up,\down}
L^\dagger_{b,\alpha}R_{b,\alpha}
L^\dagger_{b,-\alpha}R_{b,-\alpha}\ ,\nn
\rho^{(+)}_{4k_{ab}}(x)&\sim&{U}G_{ab,>}(3k_{ab})\sum_{\alpha=\up,\down}
L^\dagger_{ab,\alpha}R_{ab,\alpha}
L^\dagger_{ab,-\alpha}R_{ab,-\alpha}\ ,\nn
\rho^{(+)}_{2k_b+2k_{ab}}(x)&\sim&
\frac{U}{2}\big[G_{b,>}(k_{ab}+2k_b)+G_{b,>}(k_b+2k_{ab})\big]
\sum_{\alpha=\up,\down}\Big\{
2L^\dagger_{b,\alpha}R_{b,\alpha}
L^\dagger_{ab,-\alpha}R_{ab,-\alpha}\nn
&&+2L^\dagger_{b,\alpha}R_{ab,\alpha}
L^\dagger_{ab,-\alpha}R_{b,-\alpha}
+L^\dagger_{b,\alpha}R_{ab,\alpha}
L^\dagger_{b,-\alpha}R_{ab,-\alpha}
+L^\dagger_{ab,\alpha}R_{b,\alpha}
L^\dagger_{ab,-\alpha}R_{b,-\alpha}\Big\}\ ,\nn
\rho^{(-)}_{k_b+3k_{ab}}(x)&\sim&
\frac{U}{2}\big[3G_{b,>}(k_{ab}+2k_b)+G_{ab,>}(3k_{ab})\big]
\sum_{\alpha=\up,\down}\Big\{
L^\dagger_{ab,\alpha}R_{ab,\alpha}
L^\dagger_{ab,-\alpha}R_{b,-\alpha}
+L^\dagger_{ab,\alpha}R_{ab,\alpha}
L^\dagger_{b,-\alpha}R_{ab,-\alpha}\Big\}\ ,\nn
\rho^{(-)}_{3k_b+k_{ab}}(x)&\sim&
\frac{U}{2}\big[3G_{ab,>}(k_{ab}+2k_b)+G_{b,>}(3k_b)\big]
\sum_{\alpha=\up,\down}\Big\{
L^\dagger_{b,\alpha}R_{b,\alpha}
L^\dagger_{b,-\alpha}R_{ab,-\alpha}
+L^\dagger_{b,\alpha}R_{b,\alpha}
L^\dagger_{ab,-\alpha}R_{b,-\alpha}\Big\}\ .
\eea
These expressions can be bosonized following Ref.~\cite{GNT}, giving
\bea
\rho^{(+)}_{4k_{b}}(x)&\sim&-2UG_{b,>}(3k_{b})\ e^{i\sqrt{4\pi}\Phi_{+,c}}\
e^{i\sqrt{4\pi}\Phi_{-,c}}\ ,\nn
\rho^{(+)}_{4k_{ab}}(x)&\sim&-2UG_{ab,>}(3k_{ab})\ e^{i\sqrt{4\pi}\Phi_{+,c}}\
e^{-i\sqrt{4\pi}\Phi_{-,c}}\ ,\nn
\rho^{(+)}_{2k_b+2k_{ab}}(x)&\sim&{\cal C}_{2k_b+2k_{ab}}
\ e^{i\sqrt{4\pi}\Phi_{+,c}}\left[
\cos\big(\sqrt{4\pi}\Phi_{-,s}\big)
+\cos\big(\sqrt{4\pi}\Theta_{-,s}\big)
-\cos\big(\sqrt{4\pi}\Theta_{-,c}\big)\right]\ ,\nn
\rho^{(-)}_{3k_b+k_{ab}}(x)&\sim&{\cal C}_{3k_b+k_{ab}}
e^{i\sqrt{4\pi}\Phi_{+,c}}\left[
e^{-i\sqrt{\pi}\bar{\varphi}_{-,c}}\cos(\sqrt{\pi}\varphi_{-,s})
-e^{i\sqrt{\pi}{\varphi}_{-,c}}\cos(\sqrt{\pi}\bar{\varphi}_{-,s})
\right]\ ,\nn
\rho^{(-)}_{k_b+3k_{ab}}(x)&\sim&{\cal C}_{k_b+3k_{ab}}
e^{i\sqrt{4\pi}\Phi_{+,c}}\left[
e^{-i\sqrt{\pi}{\varphi}_{-,c}}\cos(\sqrt{\pi}\bar{\varphi}_{-,s})
-e^{i\sqrt{\pi}{\bar{\varphi}}_{-,c}}\cos(\sqrt{\pi}{\varphi}_{-,s})
\right]\ ,
\label{4kfbosons}
\eea
where ${\cal C}_p$ are non-universal prefactors that are proportional
to $U$ for small interactions and the fields $\varphi_{\pm,d}$ and
$\bar\varphi_{\pm,d}$ are chiral components of the boson field
$\Phi_{\pm,d}$ for $d=c,s$ which satisfy
\bea
\Phi_{\pm,d} &=& \varphi_{\pm,d}+\bar\varphi_{\pm,d}\ ,\nn
\Theta_{\pm,d} &=& \varphi_{\pm,d} - \bar\varphi_{\pm,d}\ .\nonumber
\eea
Once more we may refermionize Eqs.~\fr{4kfbosons} in terms 
of the new basis of bosons, i.e.
\bea
R_1+iR_2&\sim&\frac{\kappa_{+,s}}{\sqrt{\pi
    a_0}}e^{i\sqrt{4\pi}\varphi_{+,s}} ,\quad
L_1+iL_2\sim\frac{\kappa_{+,s}}{\sqrt{\pi
   a_0}}e^{-i\sqrt{4\pi}\bar{\varphi}_{+,s}}\ ,\nn
R_3+iR_4&\sim&\frac{\kappa_{-,s}}{\sqrt{\pi a_0}}
e^{i\sqrt{4\pi}\varphi_{-,s}},\quad
L_3+iL_4\sim\frac{\kappa_{-,s}}{\sqrt{\pi
    a_0}}e^{-i\sqrt{4\pi}\bar{\varphi}_{-,s}}\ ,\nn
R_5+iR_6&\sim&\frac{\kappa_{-,c}}{\sqrt{\pi a_0}}
e^{i\sqrt{4\pi}\varphi_{-,c}},\quad
L_5+iL_6\sim\frac{\kappa_{-,c}}{\sqrt{\pi
    a_0}}e^{-i\sqrt{4\pi}\bar{\varphi}_{-,c}}\ ,
\eea
where $\kappa$ are Klein factors introduced to ensure that
different Majoranas anticommute. This choice of basis for the Majorana
fermions will make the ``dictionary''~\fr{eqs:dictionary} between the 
``band'' representation and the ``chain'' representation particularly clear.
The $4k_F$ components of the charge density are local with respect to
the Majoranas
\bea
\rho^{(+)}_{4k_{b}}(x)&\propto&e^{i\sqrt{4\pi}\Phi_{+,c}}\
\left[R_5L_6-R_6L_5+i(R_5L_5+R_6L_6)\right]\ ,\nn
\rho^{(+)}_{4k_{ab}}(x)&\propto&e^{i\sqrt{4\pi}\Phi_{+,c}}\
\left[R_5L_6-R_6L_5-i(R_5L_5+R_6L_6)\right]\ ,\nn
\rho^{(+)}_{2k_b+2k_{ab}}(x)&\propto&
e^{i\sqrt{4\pi}\Phi_{+,c}}\ 
i\left[2R_4L_4+R_5L_5-R_6L_6\right]\ ,\nn
\rho^{(-)}_{3k_b+k_{ab}}(x)&\propto&
e^{i\sqrt{4\pi}\Phi_{+,c}}\left[
(L_5+iL_6)R_3-(R_5+iR_6)L_3\right]\,, \nn
\rho^{(-)}_{k_b+3k_{ab}}(x)&\propto&
e^{i\sqrt{4\pi}\Phi_{+,c}}\left[(R_5-iR_6)L_3
-(L_5-iL_6)R_3\right]\ .
\eea

\subsection{``$4k_F$'' Density Components in the Chain Picture}
\label{app:4kFchain}
In this Appendix we determine the Fourier components with momenta
close to $4k_F$ of the low-energy projections of $n_{1,l}\pm n_{2,l}$,
c.f. eqn \fr{PP}, in the chain picture. For uncoupled chains we have
\bea
n_{1,l}\pm n_{2,l}\Big|_{t_\perp=0}\propto \sum_{n\in \mathbb{Z}}
\widetilde{\rho}^{(\pm)}_{2nk_F}(x)e^{2ink_Fx}.
\eea
For non-zero $t_\perp$ this expressions gets modified to
\bea
n_{1,l}\pm n_{2,l}\propto \sum_{P\in S_\pm}
{\rho}^{(\pm)}_{P}(x)e^{iPx},
\eea
where $S_\pm$ are appropriately defined sets of momenta.
Our starting point is the bosonized expression for the $4k_F$
components of the charge density of the extended Hubbard chains
describing the uncoupled legs $\ell=1,2$ of the ladder
(i.e. $t_\perp=0=V_\perp=W_{1,2}$) 
\be
\rho_{4k_F}^{(\ell)}(x) \sim \tilde{F} e^{i4k_Fx}
e^{i\sqrt{8\pi}\Phi^{(c)}_{\ell}} + \mathrm{h.c.}\ , 
\ee
where $\tilde{F}$ is a non-universal amplitude. The sum of the $4k_F$
densities of the two legs can be expressed in terms of the rotated boson
basis \fr{fields_chain} as
\be
\widetilde{\rho}^{(+)}_{4k_F}(x)\sim
2\tilde{F}\cos(4k_Fx+\sqrt{4\pi}\Phi_{c})\cos(\sqrt{4\pi}\Phi_{f}).
\label{app_1}
\ee
We will now take into account the effects of a nonzero $t_\perp$ by
following through the same steps as in the analysis of the Hamiltonian
in section \ref{sec:chain}.
Refermionizing in terms of Majorana fermions using the identities
\bea 
\tilde\kappa e^{i\sqrt{\pi}\big(\Phi_{f}+\Theta_f\big)} 
&=& \sqrt{\pi a_0} \left(\xi_R^3 +i\eta_R\right),\nn
\tilde\kappa e^{-i\sqrt{\pi}\big(\Phi_{f}+\Theta_f\big)} 
&=& \sqrt{\pi a_0} \left(\xi_L^3+i\eta_L\right),
\eea
where $\tilde\kappa$ is a Klein factor and $\tilde\kappa^2=1$ leads to
\be
\cos(\sqrt{4\pi}\Phi_{f}) = i\pi a_0 \left(\xi_R^3 \xi_L^3 +
  \eta_R\eta_L\right). 
\ee
Finally, performing the rotation (\ref{MajoanatrafoR}) we arrive at
\bea
\cos(\sqrt{4\pi}\Phi_{f}) &\approx& i\pi a_0 \Bigg\{ \xi_R^3 \xi_L^3 + 
\frac{v_s}{v_s+v_c} 
\Big[\xi_R^1 \xi^1_L - \xi_R^2 \xi_L^2 - \frac{1}{2}e^{2iQx} 
\left(\xi_R^1-i\xi_R^2\right)\left(\xi_L^1+i\xi_L^2\right) \nn
&&\qquad\qquad\qquad-\frac{1}{2}e^{-2iQx}
\left(\xi_R^1+i\xi_R^2\right)\left(\xi_L^1-i\xi_L^2\right)\Big]\Bigg\}. 
\label{app_2}
\eea
Substituting \fr{app_2} into \fr{app_1} then gives us expressions for
the Fourier components of the total symmetric charge density of the
ladder for non-zero $t_\perp$
\bea
\rho^{(+)}_{4k_{F}}(x) &\sim& i F e^{i\sqrt{4\pi}\Phi_{c}} \left[ \xi_R^3 \xi_L^3 + \frac{v_s}{v_s+v_c}\left(\xi_R^1 \xi_L^1 -\xi_R^2 \xi_L^2\right)\right], \nn
\rho^{(+)}_{4k_F+2Q}(x) &\sim& \frac{iFv_s}{2\left(v_s+v_c\right)}e^{i\sqrt{4\pi}\Phi_{c}} 
\left(\xi_L^1+i\xi_L^2\right)\left(\xi_R^1 - i\xi_R^2\right),\nn
\rho^{(+)}_{4k_F-2Q}(x) &\sim& \frac{iFv_s}{2\left(v_s+v_c\right)}e^{i\sqrt{4\pi}\Phi_{c}} 
\left(\xi_L^1-i\xi_L^2\right)\left(\xi_R^1+i\xi_R^2\right), \nonumber
\eea
and $F=\tilde{F}\pi a_0$ is a non-universal constant. 
The analogous analysis for the antisymmetric combination of charge
densities gives the following result
\bea
\rho^{(-)}_{4k_F-Q}(x) &=& -iF  \sqrt{\frac{v_s}{2\left(v_s+v_c\right)}}e^{i\sqrt{4\pi}\Phi_{c}}\left[\left(\xi_R^1+i\xi_R^2\right)\xi_L^3+\xi_R^3 \left(\xi_L^1-i\xi_L^2\right)\right]\ ,\\
\rho^{(-)}_{4k_F+Q}(x) &=& -iF \sqrt{\frac{v_s}{2\left(v_s+v_c\right)}} e^{i\sqrt{4\pi}\Phi_{c}}\left[\left(\xi_R^1-i\xi_R^2\right)\xi_L^3+\xi_R^3\left(\xi_L^1+i\xi_L^2\right)\right]\ ,
\eea
where $F$ is the same non-universal constant as in the $(+)$ component case.

\section{Higher Harmonics of the Bond-Centered Antibonding
  Superconducting Order Parameter}  
\label{app:abP}

We consider the order parameter for bond-centered pairing in the
antibonding band: \[ \Phi_B(j) =
c_{ab,\ua}(j)c_{ab,\da}(j+1)-c_{ab,\da}(j)c_{ab,\ua}(j+1) \] and
consider the higher-order term generated by the four-fermion
interaction. We integrate out the high-energy part of the Hubbard
interaction~\fr{HubIntAct} by splitting the fermion operators into fast 
(high-energy $>$) and slow (low-energy $<$) components as shown 
in~\fr{fermiondecomp}. We separate the ``mixed'' part of the 
bond-centered pairing order parameter into four contributions
\bea
{\cal O}_1 &=& c_{ab,\uparrow,>}(x)c_{ab,\downarrow,<}(x+a_0),\qquad\qquad
{\cal O}_3 = c_{ab,\ua,<}(x)c_{ab,\da,>}(x+a_0),\nn
{\cal O}_2 &=& -c_{ab,\da,>}(x)c_{ab,\ua,<}(x+a_0),\qquad\quad~
{\cal O}_4 = -c_{ab,\da,<}(x)c_{ab,\ua,>}(x+a_0).
\label{terms}
\eea
We now discuss in some detail the perturbative averaging of the
operator ${\cal O}_1$ with respect to the interaction term $S_{\rm
  int}^{(1)}$. We have
\bea
\la {\cal O}_1S_{\rm int}^{(1)}\ra_> &=& -\frac{U}{2} \int {\rm
  d}^2{\bf y}~ \Biggl\la c_{ab,\ua,>}(x)c_{ab,\da,<}(x+a_0)\nn 
&\times&
\left[ c_{ab,\ua,<}^\dagger c_{ab,\ua,<} + c_{ab.\ua,<}^\dagger
  c_{ab,\ua,>} + c_{ab,\ua,>}^\dagger c_{ab,\ua,<} +
  c_{ab,\ua,>}^\dagger c_{ab,\ua,>} + b\leftrightarrow ab\right]({\bf
  y})\nn 
&\times&\left[ c_{ab,\da,<}^\dagger c_{ab,\da,<} + c_{ab.\da,<}^\dagger c_{ab,\da,>} + c_{ab,\da,>}^\dagger c_{ab,\da,<} + c_{ab,\da,>}^\dagger c_{ab,\da,>} + b\leftrightarrow ab\right]({\bf y}) \Biggr\ra_>. \nonumber
\eea
This can now be averaged over the high-energy parts and the resulting
expression evaluated in the continuum limit by following the same
steps as in Appendix \ref{app:4kF}. We then bosonize, following~\cite{GNT},
and the result is
\bea
\langle {\cal O}_1 S_{\rm int}^{(1)}\rangle_> \sim \frac{U}{2} \frac{\kappa_{ab,\down}\kappa_{ab,\up}}{(2\pi)^2} i e^{i\sqrt{2\pi}\Theta_{2,c}} &\Bigg\{& G_{ab,\up,-} \cos\left(\sqrt{2\pi}\Phi_{1,c}+2k_bx\right)\cos\left(\sqrt{4\pi}\Phi_{+,s}-k_{ab}a_0\right)+\nn
&& G_{ab,\up,-} \sin\left(\sqrt{2\pi}\Phi_{1,c}+2k_bx\right)\sin\left(\sqrt{4\pi}\Phi_{+,s}-k_{ab}a_0\right) +\nn
&& iG_{ab,\up,+} \sin\left(\sqrt{2\pi}\Phi_{1,c}+2k_bx\right)\cos\left(\sqrt{4\pi}\Phi_{+,s}-k_{ab}a_0\right)-\nn
&& iG_{ab,\up,+} \cos\left(\sqrt{2\pi}\Phi_{1,c}+2k_bx\right)\sin\left(\sqrt{4\pi}\Phi_{+,s}-k_{ab}a_0\right)\Bigg\}
\eea
where $G_{ab,\up,\pm} = G_{ab,\up}(2k_b-k_{ab})\pm G_{ab,\up}(k_{ab}-2k_b)$. 
There terms arise from the four-fermion products
\be
R_{ab,\down}(x+a_0) L_{ab,\up}(x) L_{b,\down}^\dagger(x)R_{b,\down}(x)\qquad\qquad
L_{ab,\down}(x+a_0) R_{ab,\up}(x) R_{b\down}^\dagger(x)L_{b,\down}(x)\ . \nonumber
\ee
These describe the coupling of ``$2k_b$'' density oscillations in the bonding band to 
bond-centered hole pairs in the antibonding band. Carrying out the
analogous analyses for ${\cal O}_2$, ${\cal O}_3$ and ${\cal O}_4$ we find
the sum of the contributions is given by
\bea
\la \Phi_B S_{\rm int}^{(1)} \ra \sim e^{i\sqrt{2\pi}\Theta_{2,c}}
&\Big[ {\cal C}_1 \cos\left(\sqrt{2\pi} \Phi_{1,c} + 2k_b
  x\right)+{\cal C}_2 \sin\left(\sqrt{2\pi}\Phi_{1,c} +
  2k_bx\right)\Big] \cos\left(\sqrt{4\pi}\Phi_{+,s}\right)+\ldots\ ,
\label{contrib1}
\eea
where the complex coefficients ${\cal C}_{1,2}$ are given in terms of
$G_{ab,\uparrow,\pm}$ and where we have retained only the terms which
contribute power-law decay to the two point function. Terms which have
zero expectation value in the $4k_b$ Mott insulator,
e.g. contributions proportional to $\sin(\sqrt{4\pi}\Phi_{+,s})$ or
$\sin(\sqrt{4\pi}\Theta_{-,s})$, have been dropped
from~\fr{contrib1}. The order parameter $\Phi_B$ being bond-centered
is important; the contributions~(\ref{contrib1}) which decay as a
power law in the $4k_b$ Mott insulating phase vanish due to
cancellation in the site-centered case. Following through the same steps
for $S_{\rm int}^{(2)}$ we find that
\bea
\la \Phi_B S_{\rm int}^{(2)}\ra&\sim& - e^{i\sqrt{2\pi}\Theta_{2,c}} \Big[ {\cal C}_3 \cos\left(\sqrt{2\pi}\Phi_{1,c}+2k_bx\right) + {\cal C}_4 \sin\left(\sqrt{2\pi}\Phi_{1,c}+2k_bx\right) \Big] \cos\left(\sqrt{4\pi}\Theta_{-,s}\right)+\ldots\ .
\label{contrib2}
\eea
Combining the two contributions gives the following result for the
interaction induced contribution to the low-energy projection of $\Phi_B(j)$
\bea
\la \Phi_BS_{\rm int}\ra \sim e^{i\sqrt{2\pi}\Theta_{2,c}}&\Bigg\{&
\left[{\cal C}_1 \cos\left(\sqrt{4\pi}\Phi_{+,s}\right) - {\cal C}_3 \cos\left(\sqrt{4\pi}\Theta_{-,s}\right)\right] \cos\left(\sqrt{2\pi}\Phi_{1,c}+2k_bx\right)+\nn
&& \left[{\cal C}_2\cos\left(\sqrt{4\pi}\Phi_{+,s}\right) - {\cal C}_4 \cos\left(\sqrt{4\pi}\Theta_{-,s}\right) \right] \sin\left(\sqrt{2\pi}\Phi_{1,c}+2k_bx\right)\Bigg\}+\ldots
\eea


\end{document}